# Temperature-dependent photoluminescence in Ge: experiment and theory
Revised 4/30/2020  15:33:00

José Menéndez, Christian D. Poweleit, and Sean E. Tilton
*Department of Physics, Arizona State University, Tempe AZ 85287-1504*
e-mail address: jose.menendez@asu.edu

We report a photoluminescence study of high-quality Ge samples at temperatures 12 K ≤ $T$ ≤ 295 K, over a spectral range that covers phonon-assisted emission from the indirect gap (between the lowest conduction band at the $L$ point of the Brillouin zone and the top of the valence band at the $\Gamma$ point), as well as direct gap emission (from the local minimum of the conduction band at the $\Gamma$ point). The spectra display a rich structure with a rapidly changing lineshape as a function of $T$. A theory is developed to account for the experimental results using analytical expressions for the contributions from LA, TO, LO, and TA phonons. Coupling of states exactly at the $\Gamma$ and $L$ points is forbidden by symmetry for the latter two phonon modes, but becomes allowed for nearby states and can be accounted for using wave-vector dependent deformation potentials. Excellent agreement is obtained between predicted and observed photoluminescence lineshapes. A decomposition of the predicted signal in terms of the different phonon contributions implies that near room temperature indirect optical absorption and emission are dominated by "forbidden" processes, and the deformation potentials for allowed processes are smaller than previously assumed.

## I. INTRODUCTION

Optical absorption and spontaneous emission are related by the so-called van Roosbroeck-Shockley (RS) equation [1-4] in semiconductors that are in thermal equilibrium with black-body radiation. A generalization of this equation to quasi-equilibrium conditions in the conduction band (CB) and in valence band (VB) leads to

$$\frac{1}{V}\frac{dR}{d\Omega d\omega} = \frac{1}{4\pi^3}\left(\frac{n_{op}\omega}{c}\right)^2 \frac{\alpha(\omega)}{\exp\left(\frac{\hbar\omega - \Delta F}{k_B T}\right) - 1}\;. \quad (1)$$

Here the left-hand side is the differential photon emission rate per unit sample volume $V$ at photon frequency $\omega$ and solid angle $\Omega$. On the right-hand side, $\alpha(\omega)$ is the absorption coefficient, $n_{op}$ is the index of refraction, $\Delta F = F_c - F_v$ the difference between the CB and VB quasi-Fermi levels, and $T$ the absolute temperature. Finally, $\hbar$, $k_B$, and $c$ denote the reduced Planck constant, Boltzmann's constant, and the speed of light in vacuum, respectively.

Equation (1) is mostly used to model photoluminescence (PL) spectra from direct gap materials, but it is well-known that its validity extends to indirect gap semiconductors [5]. Application examples include the determination of the absorption coefficient in Si devices from PL or electroluminescence measurements [6-8]. On the other hand, predictions of PL spectra from Eq. (1) using theoretical expressions for the indirect absorption are very rare, except at very low temperatures where the PL consists of sharp peaks that are rather insensitive to the detailed photon energy dependence of $\alpha(\omega)$. In the case of Ge, if we use the textbook expression for the absorption coefficient, with phonon creation and phonon annihilation components given by $\alpha^{\pm}(\omega) \propto (\hbar\omega - E_{ind} \mp \hbar\Omega_j)^2 / (E_0 - \hbar\omega)^2$, where $E_{ind}$ is the fundamental indirect gap, $E_0$ the lowest direct gap, and $\hbar\Omega_j$ a characteristic phonon energy, one obtains from Eq. (1) the PL spectra in Fig. 1.

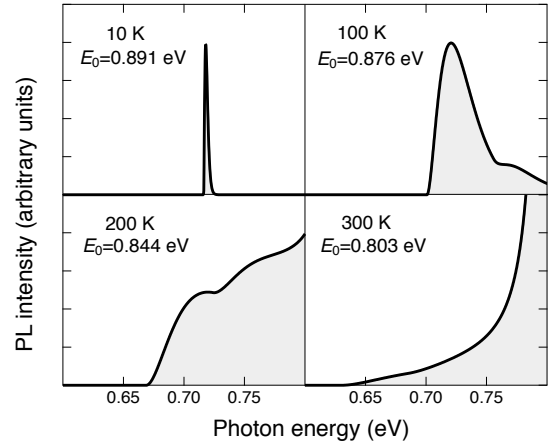

FIG. 1. Calculated indirect PL spectrum for Ge using Eq. (1) and the textbook prediction for indirect gap absorption assuming constant energy denominators in the perturbation theory expressions. The phonon energy was taken as 27.5 meV, corresponding to a LA phonon. The needed quasi-Fermi levels were computed as described in the text. The absolute and relative scales are arbitrary and were selected for visualization purposes.

It is apparent that no well-defined indirect PL peak is predicted for $T > 200$ K, in strong disagreement with experimental observations. It is therefore not surprising that researchers have resorted to more or less *ad hoc* expressions to model indirect PL from Ge [9,10]. These expressions, while useful, are unsatisfactory for the purpose of studying the electron-phonon interaction underlying indirect gap emission. On the other hand, fully *ab initio* approaches to the computation of indirect absorption spectra have become possible in recent years [11,12]. Yet the results do not lend themselves to the fitting of PL spectra via Eq. (1) nor to the modeling of the optical response of Ge-like materials and structures in the proximity of the direct gap. The latter are attracting increasing attention after the demonstration of lasing in strained Ge and in $Ge_{1-y}Sn_y$ alloys [13,14]. A rigorous yet practical theory of PL in Ge is also needed—based preferably on analytical expressions—to address suggestions that quasi-equilibrium may not be attainable in defected or highly doped Ge [15,16]. This may require a theory beyond Eq. (1) [17]. Furthermore, PL has emerged as the most reliable technique for the measurement of the separation between direct and indirect gaps in $Ge_{1-y}Sn_y$ alloys [18,19]. However, the determination of the indirect gap energy requires additional calibrations due to the lack of suitable theoretical expressions to fit the indirect gap emission.

The need for a realistic description of the indirect PL in Ge is also apparent in the field of group-IV spintronics [20], where the spin orientation of photoexcited carriers is monitored using circular polarization measurements of the emitted light [21,22]. Similarly, PL studies play an important role in the development of strained Ge microstructures and devices [23-26].

The main reason for the failure of the textbook absorption expressions in the prediction of indirect PL spectra is their strong $(E_0 - E)^{-2}$ divergence as the direct gap is approached. This divergence arises from the assumption of constant energy denominators in the second-order perturbation expressions used to compute the phonon-assisted absorption. The assumption is very good for Si but poor for Ge due to the small 0.14 eV separation between $E_{ind}$ and $E_0$. The unique challenge presented by the Ge band structure was first tackled by Hartman [27], who derived analytical expressions for the indirect absorption coefficient without the assumption of constant denominators. The Hartman expression, with suitable excitonic corrections, was recently shown to agree very well with the experimental room-temperature indirect gap absorption in Ge [28,29]. Since this expression has a weaker $(E_0 - E)^{-1/2}$ divergence, it can be expected to lead to much better PL predictions when inserted into Eq. (1).

In this paper, we present a combined experimental and theoretical study of PL in Ge. Our emphasis is not on extremely low temperatures, for which many PL studies are available [30-34], but on the intermediate range between cryogenic and room temperature, which provides the best test of theoretical predictions based on Eq. (1). Some earlier work suggested the need to include no-phonon transitions to account for indirect gap PL [9]. These transitions are very difficult to model from a microscopic perspective. To ensure that they are minimized, we carried out our experiments on the highest-quality *bulk* germanium wafers commercially available. On the other hand, Ge has a very large ambipolar diffusion coefficient [35]. This means that the PL signal in bulk samples originates from large volumes over which the photoexcited carrier concentration can vary substantially, requiring point by point calculations of the quasi-Fermi levels and reabsorption corrections to compute the signal reaching the detector. Furthermore, the carrier diffusion process has a large lateral component, so standard one-dimensional models are not suitable to compute steady-state carrier concentrations [36]. Since the required numerical solution of the realistic three-dimensional diffusion equation is impractical for the analysis of experimental data, we have developed an effective one-dimensional equation where the lateral out-diffusion appears as an effective diffusion length that depends on the laser beam waist. We find this dependence to be very significant. This indicates that the neglect of lateral out-diffusion can lead to large systematic errors in power-dependence studies.

Our results show that the PL lineshape changes dramatically with temperature, a strong indication that phonons of very different frequencies are involved in the emission process. Since the minimum of the CB is at the *L* point of the Brillouin zone (BZ), and the maximum of the valence band occurs at the Γ-point of the BZ, wave vector conservation requires that these phonons should have wave vectors near the *L* point of the BZ. Earlier low-temperature absorption and PL work has found evidence that longitudinal acoustic (LA), transverse optic (TO), longitudinal optic (LO), and transverse acoustic (TA) phonons [31,37] are involved. The frequencies are similar for the first three modes, but the TA frequencies at the *L* point are very low in tetrahedral semiconductors [38]. Therefore, the interplay between TA phonons and the higher-frequency LA, LO, and TO modes must be responsible for the observed PL lineshape changes as a function of temperature. However, the electron-phonon coupling of electronic states at the Γ and *L* points in the BZ *vanishes* for TA or LO phonons both in the CB and in the VB [31]. Accordingly, if TA- and LO-induced transitions contribute to the PL this must be due to lower-symmetry states near the Γ and *L* points. It then follows that the electron-phonon coupling (deformation potential) should be taken as a function of the wave vectors that becomes zero for states exactly at *L* and Γ. This wave vector dependence, when incorporated into the corresponding perturbation theory expressions, leads to "forbidden" absorption coefficients with a very different energy dependence compared to its allowed counterparts. We have derived some of these "forbidden" expressions using the same approximations that lead to the Hartman model for allowed LA coupling, and we will show that they play an important role in matching the PL lineshapes via Eq. (1).

The number of possible phonon-assisted transition channels becomes quite high when one includes "forbidden" electron-phonon coupling. Furthermore, if a process can occur via two or more mechanisms, there is always the possibility of interferences, so that the phase of the electron-phonon matrix elements becomes important. The spirit of our theoretical work, however, is not to include every possible process but to identify the main contributions that give a satisfactory description of the PL. We find that a single transition channel per phonon is sufficient for this purpose, leading in fact to excellent agreement with experiment. For the selection of the relevant "forbidden" processes we are guided by *ab initio* calculations of the electron-phonon interaction in Ge [39-41]. In particular, the work of Tandon *et al.* is very useful for this purpose because it presents electron-phonon matrix elements over the entire BZ [40]. In addition, there are several experimental and theoretical results that impose severe constraints on the possible coupling mechanisms. For example, the electron-phonon coupling between the $\Gamma$ and $L$ points in the CB determines the intervalley carrier scattering rate [42], the linewidth of the direct gap exciton [43], and contributes to the indirect absorption [29]. These three seemingly different phenomena have been fit with the same deformation potential $D_{LA} \sim 4.2 \times 10^{-8}$ eV/cm assuming that only LA phonons contribute. If other phonons must be included, as suggested by our PL results, then we must require that the phonon combination that fits the PL results still accounts for the other non-PL measurements. Furthermore, at the highest temperatures, clear evidence is seen for direct gap emission that is not mediated by phonons. Our theoretical model should also account for the relative strength of the direct and indirect signals. On the theoretical front, Tandon *et al.* have also calculated the electron-phonon broadening of states throughout the BZ [40], and Tyuterev *et al.* have shown from *ab initio* simulations that the contribution of TA phonons to the $\Gamma$-$L$ scattering time is negligible [39]. We will show that our PL model not only agrees self-consistently with all available experimental data, but it is also consistent with the *ab initio* predictions in Refs. [39] and [40].

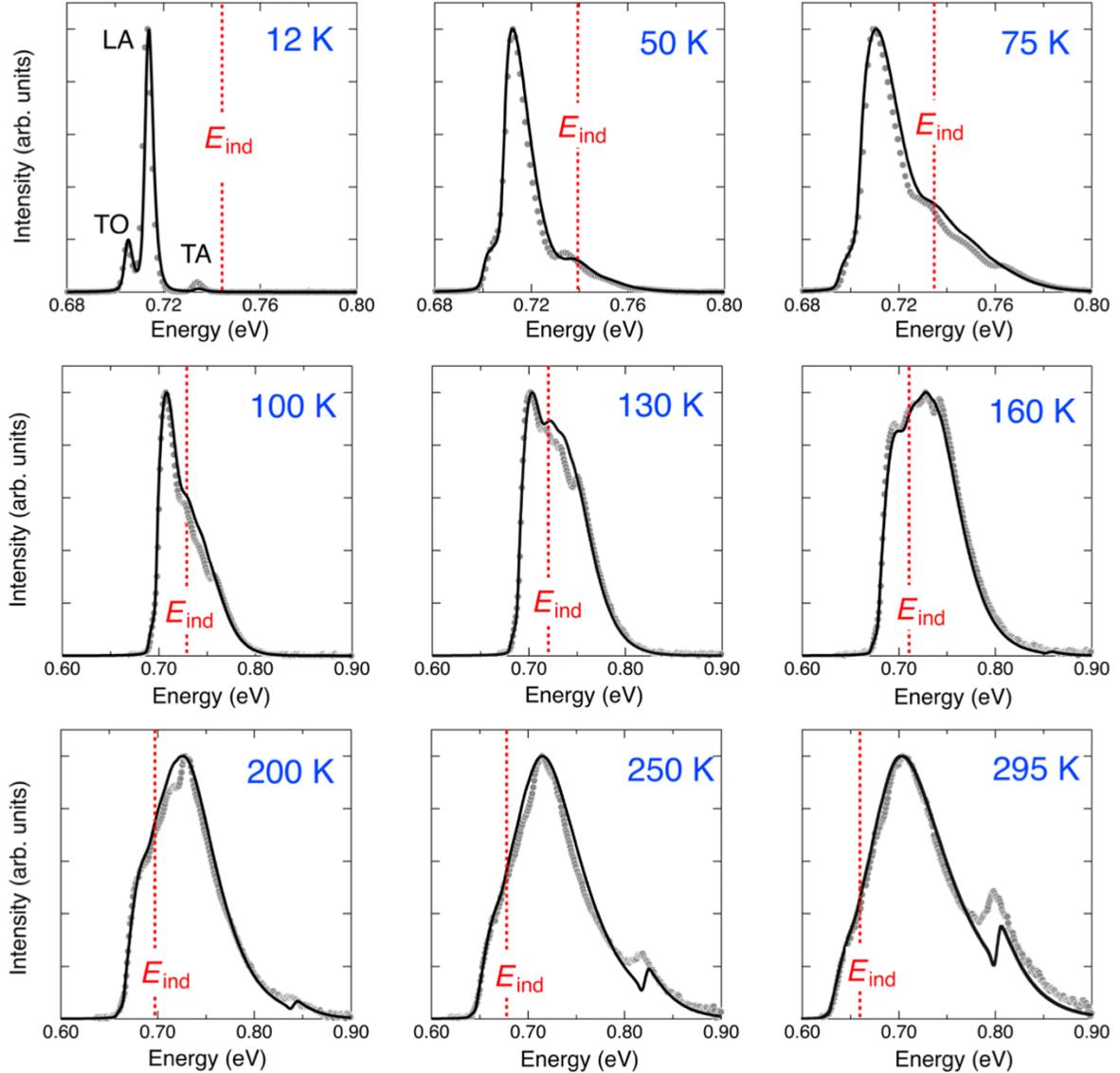

FIG. 2. Experimental (circles) and theoretically predicted (solid lines) PL spectra for Ge at nine selected temperatures. All spectra have been normalized to the same intensity for clarity. Note the expanded energy scale for the top row at the lowest temperatures. The vertical dotted lines indicate the position of the indirect gap at each temperature. Peak assignments are shown in the 12 K spectrum. The LO contribution in this spectrum appears as a weak shoulder in higher-resolution spectra [31].

## II. EXPERIMENT

The sample used for the PL experiment was an epi-ready, double-side polished Ge substrate from UMICORE [44]. The wafer's rms roughness is better than 1 nm, and the impurity concentration is below $2 \times 10^{10}$ cm$^{-3}$, corresponding to a resitivity $\varrho_e > 57$ $\Omega$cm. The sample was mounted strain-free in a CTI CRYOGENICS Model 22 refrigerator cryostat in contact with a thermal block. A single-stage refrigerator enables the block to be cooled down to 10K. The temperature is controlled via a PID feedback system with the unit's heater and temperature sensor mounted to the sample block. A 5-10-minute time lag was used to allow temperatures to stabilize after moving to the next selected temperature. Carriers were photoexcited either with a LASERGLOW TECHNOLOGIES 1064 nm (1.165 eV) laser or an EXCEL Laser Quantum 532 nm (2.33 eV) laser. Each laser system was current-controlled to adjust the output power delivered to the sample, as measured by a NEWPORT Model 1830-C power meter. The laser beam was focused with a $f = 250$ mm plano-convex lens that produces a minimum beam waist $w_0 \sim 250$ μm on the sample. The average laser power did not exceed 50 mW. The emitted light was conditioned using an 850-nm-cutoff low pass filter and a SEMROCK Raman edge filter, and focused on the entrance slit of an ACTON Spectra Pro 275 spectrometer equipped with a 600 g/mm, 1600 nm blaze grating. The dispersed light (with an energy resolution of 6 nm (5.4 meV) at 1 mm slits), was detected with an ELECTRO-OPTICAL liquid nitrogen cooled ex-InGaAs photodiode at the exit slit. The signal from the detector was processed in a single point scan lock-in amplifier (STANFORD SR830 DSP) configuration. The modulation was provided by an optical chopper that halves the net power incident on the sample. The resulting spectra were converted to an energy scale, corrected for the constant slit width during the measurements, and for the system's spectral response, which was calibrated with a tungsten lamp.

The measured PL obtained with either 1064 nm or 532 nm excitation is virtually identical, as expected if quasi-equilibrium conditions prevail in the CB and VB. Because of this strong similarity we only show here results from the 532 nm experiments. Figure 2 shows spectra for nine different temperatures between 12 K and 295 K. The spectra have been normalized to the same maximum value for better visualization of the lineshape changes. However, relative intensities are also predicted by Eq. (1) and are shown in Fig. 3. They will be discussed below.

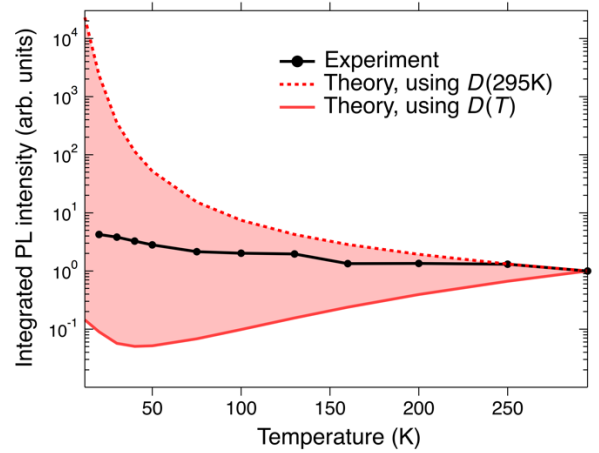

FIG. 3. Integrated Ge photoluminescence intensities as a function of temperature, normalized to unity for $T = 295$ K. The dots correspond to experimental spectra excited with 50 mW (average power) of 532 nm radiation. The solid line is the theoretical prediction using the ambipolar diffusivity from Eq. (14). The dotted line was obtained by ignoring the temperature dependence of the diffusivity and using its $T = 295$ K value.

We see that the PL line shape changes dramatically as a function of temperature, which we assign to the interplay between different phonon modes. These different contributions seem to be comparable at a temperature close to 160 K.

## III. ILLUMINATION MODEL AND QUASI-FERMI LEVELS

The quasi-Fermi levels required in Eq. (1) are obtained from the calculated steady state carrier concentrations under laser illumination. For this one solves a diffusion equation with generation and recombination terms. The usual textbook approach is to assume uniform illumination at the sample surface, which leads to a simple one-dimensional diffusion equation. However, this uniform illumination model is not good for our experimental conditions due to the large ambipolar diffusion coefficient $D$ in Ge [35]. Accordingly, one needs to solve a more complicated diffusion problem given by [36]

$$-D\left[\frac{1}{r}\frac{\partial}{\partial r}\left(r\frac{\partial \Delta n}{\partial r}\right) + \frac{\partial^2 \Delta n}{\partial z^2}\right] = G(r,z) - \frac{\Delta n}{\tau} \quad (2)$$

Here $r$ is the radial coordinate in the plane of the sample surface and $z$ is the depth coordinate. $\Delta n = \Delta p$ is the excess carrier concentration, $G$ the electron-hole generation rate per unit volume, and $\tau$ the recombination lifetime. For our experimental conditions the laser light is absorbed over distances much smaller than the depth of focus of our lens, so that we can approximate the generation rate as

$$G(r,z) = \frac{\alpha I_{\max}(1-\mathcal{R})}{\hbar \omega}\exp\left(\frac{-2r^2}{w_0^2}\right)e^{-\alpha z} \quad (3)$$

where $\alpha$ is the absorption coefficient, $I_{max}$ is the maximum intensity of the incident laser beam, $\mathcal{R}$ the sample reflectance, and $w_0$ the beam waist at the focal point.

Our PL signal obtains from an integration over $r$ and $z$, which suggests that we can compute approximate intensities by defining an average excess carrier concentration

$$\overline{\Delta n}(z) = \frac{2}{w_0^2}\int_0^{w_0} r\, dr\, \Delta n(r,z) \ . \tag{4}$$

This average carrier concentration is then used to compute the quasi-Fermi levels at depth $z$. An effective differential equation for $\overline{\Delta n}(z)$ can be derived by integrating Eq. (2) over the $r$ coordinate. We obtain

$$-\frac{2D}{w_0}\frac{\partial \Delta n}{\partial r}\bigg|_{w_0} - D\frac{\partial^2 \overline{\Delta n}}{\partial z^2} = \frac{(1-e^{-2})P_0(1-\mathcal{R})}{\pi \hbar \omega w_0^2}\alpha e^{-\alpha z} - \frac{\overline{\Delta n}(z)}{\tau} \tag{5}$$

where $P_0$ is total incident power of the laser beam. This is the standard one-dimensional equation for uniform illumination except for the first term, which we then treat in an approximate way. For this we notice that in the limit of vanishing diffusion we should obtain $\Delta n \propto \exp(-2r^2/w_0^2)$, so that $\partial \Delta n/\partial r\big|_{w_0} = -4\Delta n(w_0,z)/w_0 \simeq -\frac{4}{3}\overline{\Delta n}(z)/w_0$. Inserting this back into Eq. (5), we finally obtain a differential equation formally identical to the one-dimensional equation for uniform illumination

$$\frac{\partial^2}{\partial z^2}\overline{\Delta n}(z) - \frac{\overline{\Delta n}(z)}{L_{eff}^2} = -\frac{(1-e^{-2})P_0(1-\mathcal{R})}{\pi w_0^2 \hbar \omega D}\alpha e^{-\alpha z} \tag{6}$$

but with an effective diffusion length given by

$$\frac{1}{L_{eff}^2} = \frac{1}{L^2} + \frac{8}{3w_0^2} \tag{7}$$

where $L = \sqrt{D\tau}$. The solution is therefore [2]

$$\overline{\Delta n}(z) = \left[\frac{(1-e^{-2})P_0}{\pi w_0^2}\right] \times \left(\frac{L_{eff}^2}{D}\right)\frac{\alpha(1-\mathcal{R})}{\hbar\omega(\alpha^2 L_{eff}^2 - 1)}\left[\left(\frac{s_0 + D\alpha}{s_0 + D/L_{eff}}\right)\exp\left(-\frac{z}{L_{eff}}\right) - e^{-\alpha z}\right] \tag{8}$$

where $s_0$ is the surface recombination velocity. For our ultra-high purity sample, we estimate $\tau$ = 0.9 ms at room temperature from a model that includes Shockley-Read-Hall, Auger, and radiative recombination. Using the experimental value $D$ = 53 cm$^2$/s at 295 K ([35]) and $w_0$ = 250 μm, we find $L$ = 2,000 μm and $L_{eff}$ = 150 μm. Thus, the effect of lateral out-diffusion is dominant under our experimental conditions. With $s_0$ =100 cm/s [45], we estimate $\overline{\Delta n}(100\text{ nm})$ = 1.6×10$^{16}$ cm$^{-3}$ for $P_0$ = 100 mW from Eq. (8). But if we ignore the second term in Eq. (7), we find $\overline{\Delta n}(100\text{ nm})$ = 1.1×10$^{17}$ cm$^{-3}$, almost an order of magnitude higher. This suggests that studies of the power dependence of the PL in Ge should carefully consider the role of out-diffusion, particularly if much tighter focal waist values are used. It is interesting to compare our approximate solution to the numerical solution of the same problem for the case of InN, as computed by Cuscó et al. [36]. They find that out-diffusion reduces the peak carrier concentration by a factor of 11.2, whereas our model predicts a reduction by a factor of 12.6. Thus the agreement is very good, particularly in view of the simplicity of our approach and the availability of analytical solutions via Eq. (8).

The quasi-Fermi level $F_c$ in the CB at depth $z$ is obtained by solving the equation

$$n_i + \overline{\Delta n}(z) \simeq \overline{\Delta n}(z) = 4N_c^L(T)\left\{F_{1/2}\left(\frac{F_c - E_{ind}}{k_B T}\right) + \left(\frac{k_B T}{\Delta_L}\right)F_{3/2}\left(\frac{F_c - E_{ind}}{k_B T}\right)\right\} + N_c^\Gamma(T)\left[F_{1/2}\left(\frac{F_c - E_0}{k_B T}\right) + \frac{15}{8}\left(\frac{k_B T}{\Delta_\Gamma}\right)F_{3/2}\left(\frac{F_c - E_0}{k_B T}\right)\right] + 6N_c^\Delta(T)F_{1/2}\left(\frac{F_c - E_\Delta}{k_B T}\right) \tag{9}$$

The prefactors are given by

$$N_c^L(T) = \frac{1}{4}\left[\frac{2m_{Ld}k_B T}{\pi\hbar^2}\right]^{3/2}$$
$$N_c^\Gamma(T) = \frac{1}{4}\left[\frac{2m_{\Gamma d}k_B T}{\pi\hbar^2}\right]^{3/2} \tag{10}$$
$$N_c^\Delta(T) = \frac{1}{4}\left[\frac{2m_{\Delta d}k_B T}{\pi\hbar^2}\right]^{3/2}$$

where $m_{Ld}$, $m_{\Gamma d}$, and $m_{\Delta d}$ are the density-of-states effective masses of the $L$, $\Gamma$, and $\Delta$ valleys with energies $E_{ind}$, $E_0$ and $E_\Delta$ with respect to the top of the valence band. For the anisotropic $L$ and $\Delta$ valleys this effective mass is $(m_\perp^2 m_\parallel)^{1/3}$ where $m_\perp$ is the transverse mass and $m_\parallel$ the longitudinal mass. The functions $F_{1/2}$ and $F_{3/2}$ that appear in Eq. (9) are Fermi integrals defined as in Ref. [46]. The terms involving the $F_{3/2}$ function correspond to non-parabolicity corrections. The characteristic nonparabolicity energies $\Delta_\Gamma$ and $\Delta_L$ can be derived from k·p-theory. We obtain $\Delta_\Gamma = \frac{3}{2}[2/E_0 + 1/(E_0 + \Delta_0)]^{-1}$. The corresponding expression for $\Delta_L$ is $\Delta_L = [1/E_1 + 1/(E_1 + \Delta_1)]^{-1}$, but for higher accuracy we adjust this value to band structure calculations as described in Ref. [47].

The quasi-Femi level in the VB is obtained by solving the equation

$$n_i + \overline{\Delta n}(z) \simeq \overline{\Delta n}(z) =$$
$$N_{lh}(T)\left\{F_{1/2}\left(\frac{-F_v}{k_B T}\right) - 6.0186(k_B T)\left(\frac{3}{2}\right)F_{3/2}\left(\frac{-F_v}{k_B T}\right)\right.$$
$$\left. + 128.22(k_B T)^2\left(\frac{15}{4}\right)F_{5/2}\left(\frac{-F_v}{k_B T}\right)\right\}$$
$$+ N_{hh}(T)\left\{F_{1/2}\left(\frac{-F_v}{k_B T}\right) + 3.8263(k_B T)\left(\frac{3}{2}\right)F_{3/2}\left(\frac{-F_v}{k_B T}\right)\right.$$
$$\left. - 4.7446(k_B T)^2\left(\frac{15}{4}\right)F_{5/2}\left(\frac{-F_v}{k_B T}\right)\right\}$$
$$+ N_{so}(T) F_{1/2}\left(\frac{-\Delta_0 - F_v}{k_B T}\right) \quad (11)$$

The prefactors in this equation are given by

$$N_{hh}(T) = \frac{1}{4}\left(\frac{2 m_{hh} k_B T}{\pi \hbar^2}\right)^{3/2}$$
$$N_{lh}(T) = \frac{1}{4}\left(\frac{2 m_{lh} k_B T}{\pi \hbar^2}\right)^{3/2}, \quad (12)$$
$$N_{so}(T) = \frac{1}{4}\left(\frac{2 m_{so} k_B T}{\pi \hbar^2}\right)^{3/2}$$

where $m_{hh}, m_{lh},$ and $m_{so}$ are the effective masses of the heavy-, light-, and split-off holes, respectively. The nonparabolicity terms and their numerical prefactors were obtained from fits to the valence band density of states computed by Rodríguez-Bolívar *et al.* [48]. These fits are valid for hole energies less than 0.4 eV. The nonparabolicity corrections in Eq. (9) and (11) give an excellent account of the deviation between the measured intrinsic carrier concentrations and the predictions from a simple two-band parabolic model [49], but they could be neglected with little error for the computation of PL spectra.

The thermal occupation model can also be used to derive an expression for the Thomas-Fermi screening wave vector corresponding to the photoexcited carriers [50,51]. Neglecting the nonparabolicity components, we obtain

$$k_{TF}^2 = \frac{4\pi e^2}{k_B T} \times$$
$$\left\{4 N_c^L(T) F_{-1/2}\left(\frac{F_c - E_{\text{ind}}}{k_B T}\right) + N_c^{\Gamma}(T) F_{-1/2}\left(\frac{F_c - E_0}{k_B T}\right) \right.$$
$$\left. - \left[N_{hh}(T) + N_{lh}(T)\right] F_{-1/2}\left(\frac{-F_v}{k_B T}\right)\right\} \quad (13)$$

The ambipolar diffusion coefficient is needed at all measurement temperatures, and we compute it from the expression

$$D = \frac{D_n \mu_p + D_p \mu_n}{\mu_p + \mu_n}. \quad (14)$$

Here $\mu_n, \mu_p$ are the temperature-dependent electron and hole mobilities, respectively, from [52]. The electron and hole diffusion coefficients are related to the mobilities by the generalized Einstein relations:

$$D_n = \frac{\mu_n k_B T}{e}\left\{\frac{F_{1/2}\left[(F_c - E_{\text{ind}})/k_B T\right]}{F_{-1/2}\left[(F_c - E_{\text{ind}})/k_B T\right]}\right\}$$
$$D_p = \frac{\mu_p k_B T}{e}\left\{\frac{F_{1/2}(-F_v/k_B T)}{F_{-1/2}(-F_v/k_B T)}\right\} \quad (15)$$

where the curly brackets become unity in the non-degenerate limit. Eq. (14) gives $D(295\text{ K}) = 63.7$ cm$^2$/s and $D(135\text{ K}) = 148$ cm$^2$/s, in good agreement with the direct measurements $D(295\text{ K}) = 53$ cm$^2$/s and $D(135\text{ K}) = 142$ cm$^2$/s in Ref. [35]. However, the predicted value $D(4.2\text{ K}) = 2940$ cm$^2$/s is one order of magnitude higher than the measured excitonic diffusion coefficient $D(4.2\text{ K}) = 300$ cm$^2$/s in Ref. [53]. Furthermore, in the case of Si, Zhao [54] has found that the ambipolar diffusion coefficient follows Eq. (14) from room temperature down to 250 K, but it decreases at lower temperatures, in complete disagreement with the prediction from this equation. Efros [55] has presented a theory that attempts to explain the discrepancy in terms of carrier-carrier interactions in the photoexcited plasma, but application of this theory to Ge gives only a small correction for our typical photoexcitation levels.

## IV. ABSORPTION THEORY

### 1. Hamiltonians

The absorption processes relevant for the calculation of the photoluminescence are determined by the electron-radiation and electron-phonon interaction. In the dipole approximation the electron radiation interaction is given by

$$\mathcal{H}_{eR} = H_{eR}\left(a_\lambda^\dagger + a_\lambda\right)$$
$$H_{eR} = \left(\frac{e}{m_0}\right)\left(\frac{4\pi}{n_{op}^2 V}\right)^{1/2}\left(\frac{\hbar}{2\omega}\right)^{1/2} \sum_{\mathbf{k},m,m'}\left(\hat{\mathbf{e}}_\lambda \cdot \mathbf{P}_{m'\mathbf{k},m\mathbf{k}}\right) c_{m'\mathbf{k}}^\dagger c_{m\mathbf{k}} \quad (16)$$

where $e$ and $m_0$ are the free electron charge and rest mass, $a_\lambda^\dagger$ ($a_\lambda$) is a creation (annihilation) operator for a photon with frequency $\omega$ and polarization $\lambda$, $\hat{\mathbf{e}}_\lambda$ a unit polarization vector, and $\mathbf{P}_{m'\mathbf{k},m\mathbf{k}}$ the momentum matrix element between Bloch states $m\mathbf{k}$ and $m'\mathbf{k}$ with creation and annihilation operators $(c_{m\mathbf{k}}^\dagger, c_{m\mathbf{k}})$ and $(c_{m'\mathbf{k}}^\dagger, c_{m'\mathbf{k}})$, respectively. The electron-phonon interaction is written as

$$\mathcal{H}_{eP} = \sqrt{\frac{\hbar}{2V\rho}}$$
$$\times \sum_j \sum_{m'm} \sum_{\mathbf{k}\mathbf{q}} \sqrt{\frac{1}{\Omega_{\mathbf{q}j}}} D_{m'm}^j(\mathbf{k}+\mathbf{q},\mathbf{k})\left(a_{\mathbf{q}j} - a_{-\mathbf{q}j}^\dagger\right) c_{m'\mathbf{k}+\mathbf{q}}^\dagger c_{m\mathbf{k}} \quad (17)$$

where $D_{m'm}^j(\mathbf{k}+\mathbf{q},\mathbf{k})$ is the so-called deformation potential, and $(a_{\mathbf{q}j}^\dagger, a_{\mathbf{q}j})$ are the creation/annihilation operators for a phonon of wave vector $\mathbf{q}$ in branch $j$. Prefactors have been chosen to match the standard Conwell definitions of intervalley deformation potentials [41,56]. Theorists[57] are more likely to use the matrix element

$$g_{m'mj}(\mathbf{k},\mathbf{q}) = \sqrt{\frac{\hbar}{4\Omega_j(\mathbf{q})M}} D^j_{m'm}(\mathbf{k}+\mathbf{q},\mathbf{k}) \quad (18)$$

where $M$ is the mass of a Ge atom. Detailed *ab initio* calculations of $g_{m'mj}(\mathbf{k},\mathbf{q})$ for diamond, Si and Ge have been presented by Tandon et al. [40,58]. The deformation potential is related to the phonon eigenvectors $\vec{e}(b|\mathbf{q}j)$ by

$$D^j_{m'm}(\mathbf{k}+\mathbf{q},\mathbf{k}) = \sqrt{2}\sum_b \vec{e}(b|\mathbf{q}j)\cdot \vec{D}^b_{m'm}(\mathbf{k}+\mathbf{q},\mathbf{k}), \quad (19)$$

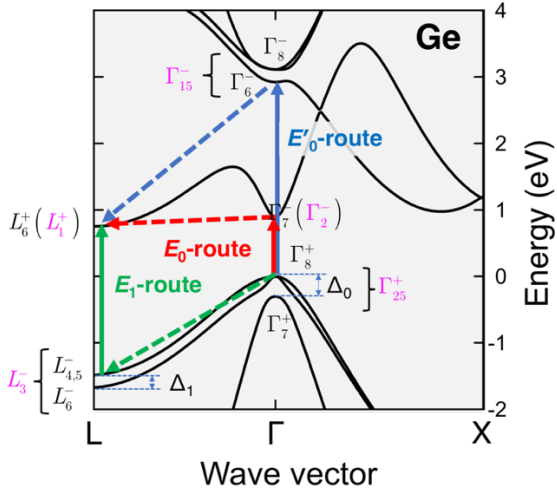

FIG. 4. Schematic depiction of the possible quantum mechanical routes to indirect and direct gap optical absorption in Ge. The solid lines indicate optical transitions induced by the electron-radiation interaction $\mathcal{H}_{eR}$; dashed-lines correspond to the electron-phonon interaction $\mathcal{H}_{eP}$. The relevant states are labeled by their double-group representations. Shown in magenta are the single group representations if spin-orbit coupling is neglected.

where $\vec{D}^b_{m'm}(\mathbf{k}+\mathbf{q},\mathbf{k})$ is a matrix element of the potential energy gradient between the Bloch states $m',\mathbf{k}+\mathbf{q}$ and $m,\mathbf{k}$ [57].

Notice that for notational compactness we omit the spin indices in Eq. (16) and (17). For all calculated interband transitions between a band $m$ and a band $m'$ below, we add over all spins. Since $\mathcal{H}_{eR}$ and $\mathcal{H}_{eP}$ do not flip spins, these spin sums introduce factors of 2.

### 2. Direct absorption

Fig. 4 shows schematically the absorption processes in Ge. Direct absorption is caused by the electron-radiation interaction only (solid red arrow in Fig. 4), so that the transition rate $R$ can be obtained using Fermi's golden rule to first order. The absorption coefficient needed for Eq. (1) is obtained as $\alpha = R_{net} n_{op}/c$, where $R_{net}$ is the *net* transition rate that accounts for recombination of the photoexcited carriers. Accordingly, we obtain

$$\alpha_{m\to m',\text{free}}(\omega) = \frac{2\pi}{\hbar}\left(\frac{n_{op}}{c}\right)\left(\frac{e}{m_0}\right)^2\left(\frac{4\pi}{n_{op}^2}\right)\left(\frac{\hbar}{2\omega}\right)\frac{1}{V}$$
$$\times \sum_{\mathbf{k}} |(\hat{\mathbf{e}}_\lambda \cdot \mathbf{P}_{m'\mathbf{k},m\mathbf{k}})|^2 [f_{m\mathbf{k}} - f_{m'\mathbf{k}}]\delta(\hbar\omega - E_{m'\mathbf{k}} + E_{m\mathbf{k}}) \quad (20)$$

for absorption from a band $m$ to a band $m'$. Here the $f$'s are the Fermi-Dirac occupation factors and the $E$'s are the energies for the Bloch states $m'\mathbf{k}$ and $m\mathbf{k}$. The evaluation of $|(\hat{\mathbf{e}}_\lambda \cdot \mathbf{P}_{m'\mathbf{k},m\mathbf{k}})|^2$ requires an average over the angle between the polarization vector and wavevector [4]. In cubic systems, this average is most easily performed by adding the values obtained from polarization directions along each of the three cartesian axes and dividing by 3. We are mainly interested in absorption across the direct gap $E_0$, for which the average is $P^2/3$, where $P$ can be obtained from the $\Gamma_7^-$ effective mass $m_e$ (equal to the density of states mass $m_{\Gamma d}$ in Eq. (10)) using $m_e^{-1} = m_0^{-1} + \frac{2}{3}(P/m_0)^2[2/E_0 + 1/(E_0+\Delta_0)]$. Here $\Delta_0$ the spin-orbit splitting at $\Gamma$. Within the parabolic approximation and under the assumption of spherical symmetry, the delta function in Eq. (20) requires that the electron energy in the CB be $E_c = s_{eH} E_0 + s_{hH}\hbar\omega$ and the valence band energy be $E_v = s_{eH}(E_0 - \hbar\omega)$, where $s_{eH} = m_e/(m_e + m_H)$, $s_{hH} = m_H/(m_e + m_H)$, and $m_H$ is the hole mass [$H$ =hh (heavy hole), *lh* (light hole)]. This implies that the square bracket containing Fermi functions can be taken out of the summation over wave vectors. The resulting expression is then proportional to the joint density of states at energy $\hbar\omega$, so that the absorption becomes

$$\alpha_{\text{free}}(\omega) = \frac{4\sqrt{2}e^2P^2}{3m_0^2\hbar^3\omega c n_{op}}\sqrt{\hbar\omega - E_0}$$
$$\times \left\{\mu_{hh}^{3/2}[f_{hh}(\hbar\omega) - f_c(\hbar\omega)] + \mu_{lh}^{3/2}[f_{lh}(\hbar\omega) - f_c(\hbar\omega)]\right\} \quad (21)$$

where $\mu_{hh}$ ($\mu_{lh}$) is the reduced effective mass for the CB electron and the heavy- (light-) hole. The "free" subscript highlights the fact that this expression does not include excitonic effects, which will be discussed below.

### 3. Indirect absorption

Indirect absorption is caused by the combined effect of the $\mathcal{H}_{eR}$ and $\mathcal{H}_{eP}$ perturbations, so that the lowest-order contributions to the transition rate can be obtained from Fermi's golden rule to *second* order. Using Eq. (16) and Eq. (17), we obtain for the *net* phonon assisted absorption from a VB band $m$ to a conduction band $m'$:

$$\alpha_{m \to m', \text{free}}^{\pm} = \frac{2\pi^2 e^2 \hbar^2}{m_0^2 \rho c n_{op}} \left(\frac{1}{\hbar\omega}\right) \frac{1}{V^2} \sum_{jkq} \left| \sum_{m''} \frac{D_{m'm''}^j(\mathbf{k+q},\mathbf{k})(\hat{\mathbf{e}}_\lambda \cdot \mathbf{P}_{m''\mathbf{k},m\mathbf{k}})}{\hbar\omega - E_{m''}(\mathbf{k}) + E_m(\mathbf{k}) - i\eta_{m''\mathbf{k},m\mathbf{k}}} + \frac{D_{m''m}^j(\mathbf{k+q},\mathbf{k})(\hat{\mathbf{e}}_\lambda \cdot \mathbf{P}_{m'\mathbf{k+q},m''\mathbf{k+q}})}{\hbar\omega - E_{m'}(\mathbf{k+q}) + E_{m''}(\mathbf{k+q}) - i\eta_{m'\mathbf{k+q},m''\mathbf{k+q}}} \right|^2$$

$$\times \frac{1}{\Omega_{qj}} \left[ \left(n_{qj} + \tfrac{1}{2} \pm \tfrac{1}{2}\right)\left(f_{m\mathbf{k}} - f_{m',\mathbf{k+q}}\right) \pm f_{m',\mathbf{k+q}}\left(1 - f_{m\mathbf{k}}\right) \right] \delta\left(E_{m'}(\mathbf{k+q}) - E_m(\mathbf{k}) - \hbar\omega \pm \hbar\Omega_{qj}\right) \quad (22)$$

where the "+" and "−" superscripts correspond to phonon creation and annihilation, respectively, and the total absorption is $\alpha_{m \to m'} = \alpha_{m \to m'}^+ + \alpha_{m \to m'}^-$. In Eq. (1) the roles are reversed, with $\alpha_{m \to m'}^+$ giving emission with the *annihilation* of a phonon, and $\alpha_{m \to m'}^-$ giving emission with the *creation* of a phonon. In Eq. (22) $n_{qj}$ is the Bose-Einstein occupation number for the phonon of branch $j$ and wave vector $\mathbf{q}$ that participates in the absorption, and the $\eta$'s represent the broadening of the intermediate states. The first term inside the square bracket correspond to electron-phonon coupling in the CB, and the second term to electron-phonon coupling in the VB. This indirect absorption expression is considerably more complicated than its direct absorption counterpart, Eq. (20), and requires additional assumptions to reduce it to analytical expressions comparable to Eq. (21). The first common approximation is to neglect the wave vector dependence of the phonon frequencies, using their value at the $L$-point. Even within this approximation, the square bracket containing occupation numbers cannot be taken out of the summation, as in the case of direct absorption. However, for the conditions in our experiments it is an excellent approximation to replace it by the expression

$$N_j^{\pm} = \left(n_j + \tfrac{1}{2} \pm \tfrac{1}{2}\right) - \left(n_j + \tfrac{1}{2} \mp \tfrac{1}{2}\right) \exp\frac{\pm\hbar\Omega_j - \hbar\omega + \Delta F}{k_B T}, \quad (23)$$

which is independent of wave vector. Here $\Omega_j$ is the frequency of the mode of branch $j$ at the wave vector of the $L$ point. Note that this approximate expression still vanishes identically at the gain threshold $\hbar\omega = \Delta F$, so that Eq. (1) does not diverge at this frequency.

Additional approximations require a careful analysis of the absorption paths in Fig. 4. For the "$E_0$-route", there is a virtual optical transition across the direct gap and a virtual phonon transition in the CB from states near $\Gamma$ to states near $L$. The "$E_0'$-route" is similar but involves a virtual optical transition across the $E_0'$-gap. For the "$E_1$-route", there is a virtual optical transition across the $E_1$ or the $E_1+\Delta_1$ gaps followed by a virtual phonon transition in the VB from states near $\Gamma$ to states near $L$. Thus the energy denominators in Eq. (22) are close to $\hbar\omega - E_0$ for the $E_0$-route, $\hbar\omega - E_0'$ for the $E_0'$-route and to $\hbar\omega - E_1$ for the $E_1$-route. But since the spectral region of maximum interest is $E_{\text{ind}} \leq \hbar\omega \leq E_0$, the $E_0$-route is enhanced by at least $\left[(E_{\text{ind}} - E_1)/(E_{\text{ind}} - E_0)\right]^2 \simeq$ 100 relative to the $E_1$-route and $\left[(E_{\text{ind}} - E_0')/(E_{\text{ind}} - E_0)\right]^2 \simeq$ 280 relative to the $E_0'$-route. Hence we can neglect all but the "$E_0$-route" if the phonon deformation potentials corresponding to the different routes are comparable in magnitude. This is the case for the LA phonon [40]. Furthermore, since we are mostly interested in the contribution from states close to $L$ and $\Gamma$, it seems reasonable to expand the deformation potential in a Taylor series around the wavevectors $\mathbf{k} = 0$ and $\mathbf{k+q} = \mathbf{k}_L = \frac{\pi}{a_0}(1,1,1)$ corresponding exactly to the indirect gap. (Here $a_0$ is the cubic lattice parameter). If the zeroth-order term $D_{m'm''}^j(\mathbf{k}_L,0)$ or $D_{m''m}^j(\mathbf{k}_L,0)$ are different from zero, the Taylor series can be truncated at this stage and one can assume a constant deformation potential. From symmetry considerations, this is the case for both LA and TO phonons [31]. On the other hand, whenever $D_{m'm''}^j(\mathbf{k}_L,0)$ and $D_{m''m}^j(\mathbf{k}_L,0)$ are zero, as is the case of TA and LO phonons [31], we must include linear or higher-order terms in the Taylor expansion of the deformation potential to obtain non-zero absorption. Such "forbidden" processes will tend to be weaker, since a deformation potential that vanishes exactly at the band extrema is not expected to be large at nearby wave vectors. However, given the experimental evidence that LO and TA phonons are involved in PL spectra, "forbidden" processes cannot be neglected. Nevertheless, it seems reasonable for these phonons to ignore the $E_1$-and $E_0'$-routes and include only the-$E_0$ route. We will make this approximation for TA and LO phonons. On the other hand, the case of TO phonons is interesting because $D_{L_6^+, \Gamma_7^-}^{\text{TO}}(\mathbf{k}_L,0) = 0$ by symmetry [31]. Thus, the $E_0$-route is forbidden for this phonon, but we cannot rule out that a "forbidden" TO process via the $E_0$ route might be stronger than the allowed $E_1$-route counterpart. However, a group-theory analysis by Thomas *et al.* [31] shows that the leading term in the Taylor expansion of the TO-phonon deformation potential is *quadratic* in the wave vector displacement from the band extrema, while the same expansion gives leading *linear* terms for TA and LO phonons. Accordingly, we will ignore "forbidden" absorption by TO phonons.

Based on all the above considerations, we propose a minimal model consisting of an allowed LA contribution via the $E_0$ route, an allowed TO-contribution via the $E_1$-route, and "forbidden" TA and LO contributions via the $E_0$ route. In all of these cases, the summation over intermediate states in Eq. (22) can be reduced to a single band. Furthermore, none of these phonons are assumed to couple simultaneously in the CB and VB, eliminating the interference implicit in Eq. (22).

### 4. Allowed TO absorption

For TO absorption via the "$E_1$-route" we can greatly simplify the calculation by assuming that the energy denominator is the same for all intermediate states and neglecting its small imaginary part. This "constant denominator approximation" is the standard textbook approach to the calculation of indirect absorption. Furthermore, since the spin-orbit splitting $\Delta_1$ is comparable to the dispersion energies that are being neglected in the constant denominator approximation, we can also disregard spin-orbit effects in the intermediate states and use an average gap $\bar{E}_1 = E_1 + \Delta_1/2$. We can then take the VB states as belonging to two doubly-degenerate $L_3^-$ single-group representations, one for each spin. There are 4×2×4=32 possible TO phonon matrix elements between these states and the light-hole/heavy-hole quartet at $\Gamma$. The calculation is tedious but straightforward using the method described by Li *et al.* [22]. Combining these matrix elements with the corresponding momentum matrix elements, carrying out the sum over intermediate states, and averaging over three light polarization directions, we arrive at

$$\alpha_{\text{TO,free}}^{\pm}(\omega) = \frac{2\pi^2 e^2 \hbar^2}{m_0^2 \rho c n_{op}}\left(\frac{1}{\hbar\omega}\right)\left(\frac{2\bar{P}^2}{3}\right)\left(\frac{4D_{\text{TO}}^2}{\Omega_{\text{TO}}}\right)\frac{N_{\text{TO}}}{\left(\hbar\omega - \bar{E}_1\right)^2} \quad (24)$$
$$\times \frac{1}{V^2}\sum_H\sum_{kq}\delta\left[E_c(k+q) - E_{vH}(k) - \hbar\omega \pm \hbar\Omega_{\text{TO}}\right]$$

The momentum matrix element $\bar{P}$ is defined so that the transverse CB effective mass at the $L$ point is given by $m_\perp^{-1} = m_0^{-1} + \left(\bar{P}/m_0\right)^2\left[1/E_1 + 1/(E_1+\Delta_1)\right]$, and $D_{\text{TO}}^2$ is the average of the squared modulus of the six different matrix elements (3 $\Gamma_{25}^+$ states and 2 $L_3^-$ states) for a TO phonon when the spin-orbit interaction is completely neglected. (The result is independent of which TO partner is used [59]). Hence $D_{\text{TO}}^2$ is directly comparable, via Eq. (18), with the results from Tandon *et al.* [40]. The sum over wave vectors can be easily converted into a double-integral over the CB and VB density of states, by first making a change of variables to $k$ and $k'$, where $k+q = k_L + k'$. Using the delta function we obtain

$$\frac{1}{V^2}\sum_H\sum_{kk'}\delta\left[E_c(k') - E_{vh}(k) - \hbar\omega \pm \hbar\Omega_{\text{TO}}\right]$$
$$= \frac{2m_\perp\left(m_\parallel\right)^{1/2}\left(m_{hh}^{3/2} + m_{lh}^{3/2}\right)}{\pi^4 \hbar^6} \quad (25)$$
$$\times \int_0^{\hbar\omega - E_{\text{ind}} \mp \hbar\Omega_{\text{TO}} - \varepsilon} d\varepsilon \sqrt{\varepsilon}\sqrt{\hbar\omega - E_{\text{ind}} \mp \hbar\Omega_{\text{TO}} - \varepsilon}$$

where we have added over all (111) valleys in Ge. Here $m_\parallel$ is the longitudinal CB mass. The integral on the rhs has a well-known analytic solution, and we finally obtain:

$$\alpha_{\text{TO,free}}^{\pm}(\omega) = \frac{4e^2}{3\pi\hbar^3 m_0 \rho c n_{op}}\left(\frac{D_{\text{TO}}^2}{\hbar\Omega_{\text{TO}}}\right)\left(\frac{\bar{P}^2}{m_0}\right)$$
$$\times m_\perp\left(m_\parallel\right)^{1/2}\left(m_{hh}^{3/2} + m_{lh}^{3/2}\right)\left(\frac{N_{\text{TO}}}{\hbar\omega}\right)\frac{\left(\hbar\omega - E_{\text{ind}} \mp \hbar\Omega_{\text{TO}}\right)^2}{\left(\hbar\omega - \bar{E}_1\right)^2} \quad (26)$$

### 5. Allowed LA absorption

LA-mediated absorption is simpler than its TO counterpart in the sense that there is a single electron-phonon matrix element connecting the $\Gamma_7^-$ and $L_6^+$ CB states. However, the absorption process follows the $E_0$ route, and we cannot use the constant denominator approximation because the dispersion energies involved are comparable in size with the denominator. If we then follow the same procedure as in the case of the TO phonons, except for the step of taking the energy denominator out of the wave vector summation, we obtain

$$\alpha_{\text{LA,free}}^{\pm}(\omega) = \frac{8e^2}{3\pi^2 \hbar^3 m_0 \rho c n_{op}}\left(\frac{D_{\text{LA}}^2}{\hbar\Omega_{\text{LA}}}\right)\left(\frac{1}{\hbar\omega}\right)\left(\frac{P^2}{m_0}\right)\frac{m_\perp\left(m_\parallel\right)^{1/2}}{\left(E_0 - \hbar\omega\right)^2}$$
$$\times N_{\text{LA}}\sum_H\left(m_H\right)^{3/2}\int_0^{\hbar\omega - E_{\text{ind}} \mp \hbar\Omega} d\varepsilon \frac{\sqrt{\varepsilon}\sqrt{\hbar\omega - E_{\text{ind}} \mp \hbar\Omega_{\text{LA}} - \varepsilon}}{\left[\frac{\hbar\omega - E_{\text{ind}} \mp \hbar\Omega_{\text{LA}} - \varepsilon}{s_{eH}\left(E_0 - \hbar\omega\right)} + 1\right]^2} \quad (27)$$

The parameter $D_{\text{LA}}^2$ is often referred to as $D_{\Gamma L}^2$ in the literature [28,29] but here we use a consistent notation for all phonon modes. Hartman[27] was the first to arrive at this integral and note that it has an analytical solution for $\hbar\omega < E_0$. The resulting absorption can be written as

$$\alpha_{\text{LA,free}}^{\pm}(\omega) = \frac{4e^2}{3\pi\hbar^3 m_0 \rho c n_{op}}\left(\frac{D_{\text{LA}}^2}{\hbar\Omega_{\text{LA}}}\right)\left(\frac{1}{\hbar\omega}\right)\left(\frac{P^2}{m_0}\right)m_\perp\left(m_\parallel\right)^{1/2} N_{\text{LA}}\sum_H s_{eH}^2\left(m_H\right)^{3/2}\left\{\frac{2\left(E_0 - \hbar\omega\right) + \left(\hbar\omega - E_{\text{ind}} \mp \hbar\Omega_{\text{LA}}\right)/s_{eH}}{\sqrt{\left(E_0 - \hbar\omega\right)}\sqrt{\left[\left(E_0 - \hbar\omega\right) + \left(\hbar\omega - E_{\text{ind}} \mp \hbar\Omega_{\text{LA}}\right)/s_{eH}\right]}} - 2\right\} \quad (28)$$

Expanding the curly bracket in Eq. (28) to second order in $x = s_{eH}^{-1}\left(\hbar\omega - E_{\text{ind}} \mp \hbar\Omega_{\text{LA}}\right)/\left(E_0 - \hbar\omega\right)$, we recover the textbook expression

$$\alpha_{\text{LA,free}}^{\pm}(\omega) = \frac{e^2}{3\pi\hbar^3 m_0 \rho c n_{op}}\left(\frac{D_{\text{LA}}^2}{\hbar\Omega_{\text{LA}}}\right)\left(\frac{P^2}{m_0}\right)$$
$$\times m_\perp\left(m_\parallel\right)^{1/2}\left(m_{hh}^{3/2} + m_{lh}^{3/2}\right) N_{\text{LA}}\left(\frac{1}{\hbar\omega}\right)\frac{\left(\hbar\omega - E_{\text{ind}} \mp \hbar\Omega_{\text{LA}}\right)^2}{\left(E_0 - \hbar\omega\right)^2} \quad (29)$$

that was used to generate Fig. 1. However, this expression is not valid for the $E_0$-route in Ge because the condition $x \ll 1$ is not satisfied except very close to the absorption edge. Most absorption studies in the past focused on precisely this absorption edge spectral region, which may explain why Eq. (28) was not compared with experiment until very recently [28,29]. For the range of temperatures of our PL experiments, however, the absorption is needed over a broad spectral range and Eq. (28) must be used. The most important difference between the two expressions is that the $(E_0 - \hbar\omega)^{-2}$ divergence in Eq. (29) is replaced by a weaker $(E_0 - \hbar\omega)^{-1/2}$ divergence in Eq. (28). This should have a dramatic impact on the predicted PL lineshape and likely eliminate the discrepancy with experiment schematically illustrated in Fig. 1.

## 6. "Forbidden" LO and TA absorption

The Taylor expansion of the deformation potential $D_{cc}^j(\mathbf{k}_L + \mathbf{k}', \mathbf{k})$ connecting states near the $\Gamma$ point with states near the $L$ point in the lowest CB gives linear terms proportional to the components of the vectors $\mathbf{k}'$ and $\mathbf{k}$. We will refer to the terms linear in $\mathbf{k}'$ as "near-$L/\Gamma$ terms" and those linear in $\mathbf{k}$ as "near-$\Gamma/L$ terms" There is an important difference between them. For $\hbar\omega \to E_0$, $\mathbf{k} \to 0$, so that the electron-phonon coupling via near-$\Gamma/L$ terms will vanish. This should further suppress the $(E_0 - \hbar\omega)^{-1/2}$ divergence that appears for allowed processes via the $E_0$-route. On the other hand, near-$L/\Gamma$ terms should not suppress the divergence, and therefore they might be expected *a priori* to represent the strongest contribution. A group theory analysis by Thomas *et al.* [31] indicates that these terms are:

$$D_{cc}^{\text{LO}}(\mathbf{k}_L + \mathbf{k}', \mathbf{k}) \simeq d'_{\text{LO}} k'_z$$
$$D_{cc}^{\text{TA}}(\mathbf{k}_L + \mathbf{k}', \mathbf{k}) \simeq d'_{\text{TA}}(\mathbf{k}'_\rho \cdot \hat{\mathbf{e}}_{\text{TA}})$$ . (30)

where $k'_z$ and $k'_\rho$ are the longitudinal and transverse vector components of $\mathbf{k}'$ and $\hat{\mathbf{e}}_{\text{TA}}$ is a unit polarization vector for the TA modes, which except for a trivial phase factor can be chosen as either one of the phonon eigenvectors over the two-atom unit cell.

We start first with the LO phonon. The derivation of the absorption expression is very similar to the LA phonon case, except that in the wave vector summation in Eq. (22) there is an extra factor of $k'_z$ due to Eq. (30). When the expression is finally converted to an integral over the energy, we end up with

$$\alpha_{\text{LO,free}}^{\pm}(\omega) = \frac{16 e^2}{3\pi^2 \hbar^3 m_0 \rho c n_{op}} \left(\frac{m_\| d'^2_{\text{LO}}}{3\hbar^3 \Omega_{\text{LO}}}\right)\left(\frac{1}{\hbar\omega}\right)\left(\frac{P^2}{m_0}\right)\frac{m_\perp (m_\|)^{1/2}}{(E_0 - \hbar\omega)^2}$$
$$\times N_{\text{LO}} \sum_H m_H^{3/2} \int_0^{\hbar\omega - E_{\text{ind}} \mp \hbar\Omega_{\text{LO}}} d\varepsilon \frac{\varepsilon^{3/2} \sqrt{\hbar\omega - E_{\text{ind}} \mp \hbar\Omega_{\text{LO}} - \varepsilon}}{\left[\frac{\hbar\omega - E_{\text{ind}} \mp \hbar\Omega_{\text{LO}} - \varepsilon}{s_{eH}(E_0 - \hbar\omega)} + 1\right]^2}$$ (31)

The definite integral in Eq. (31) can also be given in an analytically closed form for $\hbar\omega < E_0$. We then obtain for the absorption

$$\alpha_{\text{LO,free}}^{\pm}(\omega) = \frac{8 e^2}{3\pi \hbar^3 m_0 \rho c n_{op}} \left(\frac{m_\| d'^2_{\text{LO}}}{3\hbar^3 \Omega_{\text{LO}}}\right)\left(\frac{1}{\hbar\omega}\right)\left(\frac{P^2}{m_0}\right) m_\perp (m_\|)^{1/2} N_{\text{LO}}$$
$$\times \sum_h \frac{s_{eH}^2 (m_H)^{3/2}}{\sqrt{(E_0 - \hbar\omega)}} \left\{ (\hbar\omega - E_{\text{ind}} \mp \hbar\Omega_{\text{LO}})\left(-3\sqrt{(E_0 - \hbar\omega)} + \sqrt{(E_0 - \hbar\omega) + (\hbar\omega - E_{\text{ind}} \mp \hbar\Omega_{\text{LO}})/s_{eH}}\right)\right.$$ (32)
$$\left. + 4 s_{eH}(E_0 - \hbar\omega)\left(-\sqrt{(E_0 - \hbar\omega)} + \sqrt{(E_0 - \hbar\omega) + (\hbar\omega - E_{\text{ind}} \mp \hbar\Omega_{\text{LO}})/s_{eH}}\right)\right\}$$

Expanding Eq. (32) to third order in $x = s_{eH}^{-1}(\hbar\omega - E_{\text{ind}} \mp \hbar\Omega_{\text{LO}})/(E_0 - \hbar\omega)$ we obtain

$$\alpha_{\text{LO,free}}^{\pm}(\omega) = \frac{e^2}{3\pi \hbar^3 m_0 \rho c n_{op}} \left(\frac{m_\| d'^2_{\text{LO}}}{3\hbar^3 \Omega_{\text{LO}}}\right)\left(\frac{P^2}{m_0}\right)$$
$$\times m_\perp (m_\|)^{1/2} (m_{hh}^{3/2} + m_{lh}^{3/2})\left(\frac{N_{\text{LO}}}{\hbar\omega}\right) \frac{(\hbar\omega - E_{\text{ind}} \mp \hbar\Omega_{\text{LO}})^3}{(E_0 - \hbar\omega)^2}$$ , (33)

which has the cubic dependence on the shift from the absorption edge that has been anticipated for "forbidden" absorption [37]. As in the case of allowed absorption, however, this expression is not valid for the $E_0$-route in Ge, and we must use Eq. (32).

For TA phonons, the dot product in Eq. (30) introduces an extra angular factor that complicates the calculation. If we simply average over the angle prior to inserting into the absorption expression, we generate an extra factor of ½ that is canceled by the summation over the two TA modes. Within this approximation, the expression for the absorption mediated by TA phonons is the same as Eq. (32) with the substitution $m_\| d'^2_{\text{LO}} \to 2 m_\perp d'^2_{\text{TA}}$ in the first bracket and the obvious use of TA frequencies instead of LO frequencies throughout. An inspection of the Tandon *et al.* [40] calculations, however, shows that $d'_{\text{TA}}$ is considerably smaller than $d'_{\text{LO}}$. Furthermore, a TA absorption mechanism along the lines of Eq. (32) implies that TA phonons participate in the relaxation of electrons from the $\Gamma$- local minimum of CB to the $L$-valley, and in the broadening of the direct gap exciton. But *ab initio* simulations of carrier relaxation show no TA involvement [39], and Li *et al.* were able to explain the exciton broadening quantitatively based only on LA phonon coupling [43,60]. These observations suggest that the TA "forbidden" absorption may be dominated by near-$\Gamma$/$L$ processes corresponding to those terms in the expansion of $D_{cc}^j(\mathbf{k}_L + \mathbf{k}', \mathbf{k})$ that are proportional to the components of $\mathbf{k}$. Indeed, the corresponding linear coefficient $d_{TA}$ along the (110) direction satisfies $d_{TA} \gg d'_{TA}$ according to Tandon *et al.* [40]. A rigorous calculation of this "forbidden" absorption is complicated because the anisotropy of $d_{TA}$ is not compatible with the spherical symmetry of the near-$\Gamma$ states. Therefore, we simply assume that the deformation potential is given by $d_{TA}k$, where $k$ is the magnitude of $\mathbf{k}$. Carrying out the calculation with the extra factor of $k$ in Eq. (22), we arrive at

$$\alpha_{TA,free}^{\pm}(\omega) = \frac{16e^2}{3\pi^2\hbar^3 m_0 \rho c n_{op}} \left(\frac{d_{TA}^2}{\hbar^3 \Omega_{TA}}\right)\left(\frac{1}{\hbar\omega}\right)\left(\frac{P^2}{m_0}\right) m_\perp \sqrt{m_\parallel} N_{TA}$$
$$\times \sum_H \frac{(m_H)^{5/2}}{(\hbar\omega - E_0)^2} \int_0^{\hbar\omega \mp \hbar\Omega_{TA} - E_{ind}} d\varepsilon \frac{\varepsilon^{3/2}\sqrt{(\hbar\omega \mp \hbar\Omega_{TA} - E_{ind} - \varepsilon)}}{\left(\frac{\varepsilon}{s_{eH}(E_0 - \hbar\omega)} + 1\right)^2} \quad (34)$$

The definite integral in Eq. (34) can also be calculated exactly for $\hbar\omega < E_0$. We then obtain for the absorption

$$\alpha_{TA,free}^{\pm}(\omega) = \frac{8e^2}{3\pi\hbar^3 c n_{op}\rho m_0}\left(\frac{d_{TA}^2}{\hbar^3\Omega_{TA}}\right)\left(\frac{1}{\hbar\omega}\right)\left(\frac{P^2}{m_0}\right) m_\perp\sqrt{m_\parallel} N_{TA}^\pm \sum_H (m_H)^{5/2} s_{eH}^2$$
$$\times \frac{(4s_{eH}(E_0-\hbar\omega) + \hbar\omega \mp \hbar\Omega_{TA} - E_{ind})\left(\sqrt{s_{eH}(E_0-\hbar\omega) + \hbar\omega \mp \hbar\Omega_{TA} - E_{ind}} - \sqrt{s_{eH}(E_0-\hbar\omega)}\right) - 2(\hbar\omega \mp \hbar\Omega_{TA} - E_{ind})\sqrt{s_{eH}(E_0-\hbar\omega)}}{\sqrt{s_{eH}(E_0-\hbar\omega) + \hbar\omega \mp \hbar\Omega_{TA} - E_{ind}}} \quad (35)$$

This expression should contain an additional factor of 2 due to the degeneracy of the TA branches. We chose not to include it for better comparison of $d_{TA}$ with Ref. 40, where it is shown that one the TA branches has negligible coupling. While Eq. (35) is clearly as invalid for $\hbar\omega > E_0$ as Eqs. (28) and (32), it does not diverge as the direct gap is approached. Thus, the anticipated further suppression of the divergence in this type of "forbidden" absorption is confirmed. This will have a profound impact on our fits of the PL lineshapes. It will not escape the reader that, since the deformation potential is squared in Eq. (22), the expression for "forbidden" absorption will contain mixed terms of the form $kk'$. The contribution from such terms can also be calculated. Not surprisingly, they yield a weak logarithmic divergence as $\hbar\omega \to E_0$. We will ignore these contributions but they can be incorporated in future refinements. Carrying out the same expansion that leads to Eq. (33), we find that the constant-denominator limit of Eq. (35) is

$$\alpha_{TA,free}^\pm(\omega) = \frac{e^2}{3\hbar^3 m_0 n_{op} \pi c \rho}\left(\frac{d_{TA}^2}{\hbar^3\Omega_{TA}}\right)\left(\frac{1}{\hbar\omega}\right)\left(\frac{P^2}{m_0}\right) N_{TA}^\pm$$
$$\times m_\perp\sqrt{m_\parallel}\sum_h (m_h)^{5/2}\frac{(\hbar\omega \mp \hbar\Omega_{TA} - E_{ind})^3}{(\hbar\omega - E_0)^2}, \quad (36)$$

which has the same photon-frequency dependence as Eq. (33).

## 7. Broadening

While we have succeeded at developing analytical expressions for the main phonon channels that contribute to the indirect gap PL in Ge, we have found that in the case of the $E_0$-route these expressions are not valid $\hbar\omega > E_0$, and some of them diverge as $\hbar\omega \to E_0$. This is a matter of concern because we are interested in calculating the PL over an emission range below and above $E_0$. While the $\hbar\omega > E_0$ range is less important for indirect PL (because the emitted light in this range is mostly reabsorbed, and it overlaps with the direct gap emission), it is extremely difficult to carry out any numerical analysis of the data with expressions that contain mathematical singularities. Furthermore, the ability to reproduce the experimental relative intensity of direct- and indirect-gap PL is an important test of theory, and this requires that the calculated indirect-gap PL at the direct gap energies be at least mathematically meaningful. The solution to this problem is to restore the broadening parameter in Eq. (22), which was set equal to zero in order to obtain analytical expressions for the absorption. Adding the neglected imaginary part to the integrals in Eq. (27), (31), and (34) requires that they be computed numerically. If we proceed this way, however, we find that the magnitude of the absorption at $\hbar\omega > E_0$ depends strongly on the broadening parameter. For example, in the case of allowed LA absorption, the absorption strength has a $1/\eta$ dependence for $\hbar\omega > E_0$. Therefore, we need a realistic theory of intermediate state broadening.

For states at the $\Gamma$-point of the BZ, Li *et al.* have studied their broadening experimentally by monitoring the width of the direct gap exciton as a function of hydrostatic pressure [43]. Other experimental studies involve the time-resolved dynamics of electrons in the CB [42,61-63]. *Ab initio* theoretical methods have been used to model such experiments [39,41,64,65], and in the case of Tandon *et al.* to compute the electron-phonon broadening of the electronic states in Si and Ge over the entire BZ [40]. The broadening is related to the transition rate $R$ by $\eta = R\hbar/2$. Using Fermi's golden rule, we then obtain

$$\eta_{mk} = \left(\frac{\pi\hbar}{2\rho V}\right) \sum_{m'qj} \frac{1}{\Omega_{qj}} \left|D_{m'm}^{j}(\mathbf{k}+\mathbf{q},\mathbf{k})\right|^{2}$$
$$\times \{(n_{qj}+1)\delta[E_{m'}(\mathbf{k}+\mathbf{q})-E_{m}(\mathbf{k})+\hbar\Omega_{qj}] \quad (37)$$
$$+ n_{qj}\delta[E_{m'}(\mathbf{k}+\mathbf{q})-E_{m}(\mathbf{k})-\hbar\Omega_{qj}]\}$$

where we have used $\Omega_{qj} = \Omega_{-qj}$. The predicted broadening $\eta_c(\varepsilon)$ of a CB state at an energy $\varepsilon$ above the minimum of the CB at Γ can then be calculated "self-consistently" with the above absorption calculations by including the same electron-phonon mechanisms. The calculation is straightforward using the same approximations and conversions to density-of-states integrations, and we obtain $\eta_c(\varepsilon) = \eta_c^+(\varepsilon) + \eta_c^-(\varepsilon)$, with

$$\eta_c^{\pm}(\varepsilon) =$$
$$\frac{\sqrt{2}}{\pi\rho}\left(\frac{D_{LA}^2}{\hbar^2\Omega_{LA}}\right)m_\perp\sqrt{m_\parallel}\left(n_{LA}+\tfrac{1}{2}\pm\tfrac{1}{2}\right)\sqrt{E_0+\varepsilon-E_{ind}\mp\hbar\Omega_{LA}}$$
$$+\frac{2\sqrt{2}}{3\pi\rho}\left(\frac{d_{LO}'^2}{\hbar^4\Omega_{LO}}\right)m_\perp m_\parallel^{3/2}\left(n_{LO}+\tfrac{1}{2}\pm\tfrac{1}{2}\right)\left(E_0+\varepsilon-E_{ind}\mp\hbar\Omega_{LO}\right)^{3/2}$$
$$+\frac{4\sqrt{2}}{3\pi\rho}\left(\frac{d_{TA}'^2}{\hbar^4\Omega_{TA}}\right)m_\perp^2\sqrt{m_\parallel}\left(n_{TA}+\tfrac{1}{2}\pm\tfrac{1}{2}\right)\left(E_0+\varepsilon-E_{ind}\mp\hbar\Omega_{TA}\right)^{3/2} \quad (38)$$
$$+\frac{2\sqrt{2}}{\pi\rho}\left(\frac{d_{TA}^2}{\hbar^4\Omega_{TA}}\right)m_c m_\perp\sqrt{m_\parallel}\left(n_{TA}+\tfrac{1}{2}\pm\tfrac{1}{2}\right)\varepsilon\sqrt{E_0+\varepsilon-E_{ind}\mp\hbar\Omega}$$
$$+\frac{2}{\sqrt{2}\pi\rho}\left(\frac{D_{\Delta\Gamma}^2}{\hbar^2\Omega_{\Delta}}\right)m_\Delta^{3/2}\left(n_\Delta+\tfrac{1}{2}\pm\tfrac{1}{2}\right)\sqrt{E_0+\varepsilon-E_\Delta\mp\hbar\Omega_\Delta}$$

where it is understood that the contribution from each term is zero when the arguments in the factors containing radicals become negative. The first term is identical to the LA-phonon contribution considered by Li *et al*. [43], while the next three terms correspond to LO and TA coupling in the CB, as discussed in the absorption calculations. The last term was introduced by Li *et al*. [43] and corresponds to electron-phonon scattering to the third lowest valley in the CB, along the (Δ,0,0) direction. The broadening needed in Eq. (22) depends on the CB and VB states connected by the momentum operator, but since the top of the valence band has zero broadening and the energy of the dominant heavy-hole states is small compared with the energy of the associated electron states, we can ignore the valence band and assume that the broadening is given by Eq. (38) alone.

Inserting Eq. (38) in the expressions containing the integrals over the energy, the absorption can be computed numerically. While one-dimensional integrals are very fast in a modern PC and fitting routines incorporating such integrals run in a reasonable time, in the case of the PL calculations the number of such integrals is very high due to the need for depth sampling. For this reason, we have generated *ad hoc* analytical expressions that match the numerical results for typical Ge parameters. For $\hbar\omega \leq E_0$, we simply add $\eta_c(0)/2$ to the energy $E_0$ in Eqs. (28), (32), and (35), where $\eta_c(0)$ is computed from Eq. (38). For $\hbar\omega > E_0$ we write $\alpha_{H,j}^{\pm}(\omega) = \alpha_{H,j}^{\pm}(E_0/\hbar)G_{H,j}^{\pm}(\omega)$, where $\alpha_{H,j}^{\pm}(\omega)$ is the heavy- or light-hole component of the absorption assisted by phonon $j$. The functions $G^{\pm}(\omega)$ are selected by noting that the corresponding integrals can be viewed as the product of a function times a Lorentzian, which can be roughly approximated as proportional the product of the value of the function at the peak of the Lorentzian times the width of the Lorentzian. Starting with such trial functions, we find by inspection of the numerical results that

$$G_{H,LA}^{\pm}(\omega) = \sqrt{\frac{\hbar\omega - E_{ind} \mp \hbar\Omega_{LA}}{E_0 - E_{ind} \mp \hbar\Omega_{LA}}}\sqrt{\frac{(\hbar\omega - E_0) + \eta_c[s_{hH}(\hbar\omega - E_0)]}{\eta_c[s_{hH}(\hbar\omega - E_0)]}}$$
$$G_{H,i}^{\pm}(\omega) = \left(\frac{\hbar\omega - E_{ind} \mp \hbar\Omega_i}{E_0 - E_{ind} \mp \hbar\Omega_i}\right)\sqrt{\frac{(\hbar\omega - E_0) + \eta_c[s_{hH}(\hbar\omega - E_0)]}{\eta_c[s_{hH}(\hbar\omega - E_0)]}} \quad (39)$$

where $i = $ TA,LO in the second equation. A comparison of the absorption computed with the modified analytical expressions and the numerical integrations shows that the agreement is nearly perfect below $E_0$ and for tens of meV above it, and within 20% at energies 0.2 eV above $E_0$. This is adequate for our purposes.

## 8. Excitonic effects

The absorption theory we have developed so far ignores excitonic effects. These effects are dominant at very low temperatures, but even at room temperature they have been shown to be critical to bring theory and experiment into agreement, both for the direct [66,67] and indirect absorption in Ge [28,29]. In PL experiments, however, the photoexcited carrier concentration can be orders of magnitude higher than typical photoexcitation levels in absorption measurements, and the associated screening of the excitonic interaction cannot be neglected *a priori*.

### 1. Direct gap excitons

A simple but accurate way to treat the effect of photoexcited carriers is to use the Hulthen potential to model screened excitons [68]. This potential mimics the Yukawa-like expression for the screened Coulomb interaction, with the significant benefit that analytical solutions are known. Furthermore, Tanguy [69] has found a surprisingly simple analytical form for the complex dielectric function at a direct gap modified by a Hulthen exciton. Applied to the direct gap of Ge, this theory amounts to rewriting Eq. (21) as

$$\alpha(\omega) = \frac{4\sqrt{2}e^2P^2\omega}{3m_0^2\hbar cn_{op}}\{\mu_{hh}^{3/2}\mathcal{H}_2(\omega,\Gamma,R_{y,hh},g)[f_{hh}(\hbar\omega) - f_c(\hbar\omega)]$$
$$+ \mu_{lh}^{3/2}\mathcal{H}_2(\omega,\Gamma,R_{y,lh},g)[f_{lh}(\hbar\omega) - f_c(\hbar\omega)]\} \quad (40)$$

where $\mathcal{H}_2(\omega,\Gamma,R,g)$ is the imaginary part of

$$\mathcal{H}(\omega,\Gamma,R,g) = \frac{\sqrt{R}}{(\hbar\omega+i\Gamma)^2} \times$$
$$[\tilde{g}(\xi(\hbar\omega+i\Gamma)) + \tilde{g}(\xi(-\hbar\omega-i\Gamma)) - 2\tilde{g}(\xi(0))] \quad (41)$$

Here $\Gamma$ is the Lorentzian broadening of the exciton and $R = \mu e^4/(2\hbar^2\varepsilon_0^2)$ is the exciton Rydberg energy. In this expression $\varepsilon_0$ is the static dielectric constant. The functions $\tilde{g}(\xi)$ and $\xi(z)$ are defined as

$$\xi(z) = \frac{2}{\left(\frac{E_0-z}{R}\right)^{1/2} + \left(\frac{E_0-z}{R}+\frac{4}{g}\right)^{1/2}} \quad (42)$$

and

$$\tilde{g}(\xi) = -2\psi\left(\frac{g}{\xi}\right) - \frac{\xi}{g} - 2\psi(1-\xi) - \frac{1}{\xi} \quad (43)$$

where $\psi(z)$ is the Digamma function. The parameter $g$ characterizes the screening. Perfect screening obtains for $g \to 0$, for which Eq. (40) reduces to Eq. (21) in the limit of vanishing Lorentzian broadening. Equation (40) implies that the absorption consists of a heavy-hole exciton plus a light-hole exciton. This additivity is not entirely justified, because the excitonic interaction is not diagonal in these states. Therefore, one must—in principle—solve a more complicated three-band problem. However, as discussed in [29], the error made by adding separate heavy- and light-direct gap excitons is small. Furthermore, we have applied such a model to strained GaAs [70], where one sees distinct light- and heavy-hole excitonic peaks, and we find excellent agreement with their relative strengths. We have also improved our model by including non-parabolicity and the wave vector dependence of the momentum matrix elements. These effects, however, are less important for PL calculations and will be discussed elsewhere.

For $g \to \infty$, the Hulthen potential approaches the bare Coulomb potential. At intermediate values, $g = 1$ determines the onset of bound states and corresponds to the so-called excitonic Mott criterion. By numerically comparing the Hulthen potential with the Yukawa potential, Bányai and Koch [71] find

$$g = \frac{12}{\pi^2 a_B k_{TF}} \quad (44)$$

where $a_B = \hbar^2\varepsilon_0/(\mu e^2)$ is the Bohr radius and $k_{TF}$ the Thomas-Fermi screening wave vector. Using Eq. (13) it is then possible to compute $g$, and we show in Fig. 5 results for different photoexcitation levels.

We find that the sharp peak corresponding to bound excitons disappears at relatively low densities in the $10^{16}$ cm$^{-3}$ range. This is in very good agreement with experimental results and more rigorous theoretical calculations in Ref. [73]. However, excitonic enhancements in the continuum remain significant at much higher photoexcitation levels and must be included in a realistic description of the direct-gap

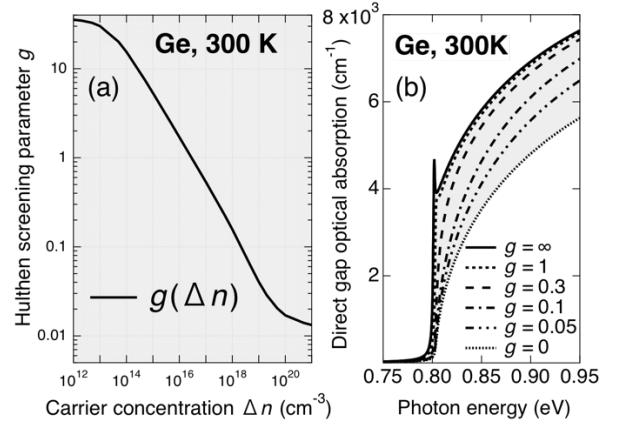

FIG. 5. (a) Calculated screening parameter $g$ in the Hulthen potential as a function of the photoexcited carrier concentration. (b) Absorption coefficient for $E_0 = 0.803$ eV, $\Gamma = 1$ meV, and other Ge parameters from Table I. Notice that the bound exciton sharp peak disappears for a Mott density $\Delta n \sim 3\times 10^{16}$, but the excitonic enhancement in the continuum persists to much higher photoexcitation levels. The bound exciton peak is not observed experimentally at 300K because the broadening parameter is at least twice the value used in this simulation [72].

PL. In particular, notice that the continuum excitonic enhancement increases the slope of the absorption curve just above the band gap relative to the perfect screening case. This affects the lineshapes computed with Eq. (1).

### 2. Indirect gap excitons

The observation that excitonic effects are important in direct gap PL experiments suggests that the same should be true for indirect gap emission. In fact, recent work has shown that the incorporation of excitonic effects is crucial to match the value of $D_{LA}$ obtained from absorption measurements with the value deduced from time resolved measurements of carrier dynamics [28,29]. Since the effect of screening is modest for our experimental conditions, we use the full Coulomb interaction for our estimates. Analytical expressions for indirect excitonic absorption were derived in a classic paper by Elliott [74] under the constant denominator assumption, which, as discussed above, is not valid for the $E_0$ route in Ge. An excitonic theory applicable to Ge has only recently been developed [28,29], but it involves numerical triple integrals for each photon frequency. The approach is therefore impractical for the computation of PL spectra, which requires the calculation of absorption at all sampling depths. We will therefore use Elliott's constant denominator theory to compute excitonic effects. The major differences between the excitonic enhancement computed from Elliott's theory and the theory of Refs. [28,29] is that the former overestimates somewhat the enhancement at the onset of absorption and underestimates it when the photon energy approaches the direct gap. But at intermediate energies corresponding to the maxima of the indirect PL at all but the lowest temperatures, the two theories give similar enhancements.

Aside from the constant denominator approximation, a second issue affecting excitonic calculations is that analytical expressions can only be obtained under the assumption of spherical symmetry. This is of course not valid in Ge, given the strong anisotropy of its lowest CB valley. However, a spherical model can be justified as a first approximation by noting that the excitonic Hamiltonian can be written as a sum whose first term does posses spherical symmetry, as shown by Altarelli and Lipari (A-L) [75,76]. The spherical component of the A-L Hamiltonian corresponds to effective electron and hole masses

$$\begin{aligned} m_L &= 3m_\parallel m_\perp/(2m_\parallel + m_\perp) \\ m_h &= 2m_{lh}m_{hh}/(m_{lh} + m_{hh}) \end{aligned} \quad (45)$$

This exciton is doubly-degenerate (since it maps into the the heavy-hole and light-hole transitions), and its binding energy is $R_y = e^4\mu_{Lh}/(2\hbar^2\varepsilon_0^2)$, with $\mu_{Lh}^{-1} = m_L^{-1} + m_h^{-1}$. The excitonic translational mass is $M_{Lh} = m_L + m_h$. Using values from Table I, we obtain $R_y = 2.65$ meV and $M_{Lh} = 0.195$. On the other hand, the full A-L Hamiltonian yields a lower and upper exciton, with binding energies of 4.20 meV and 3.18 meV, respectively, and anisotropic translational masses that are different for each exciton and strongly non-parabolic [75]. It seems then reasonable to "renormalize" the masses in Eq. (45) so that they reproduce a suitable average of the two A-L binding energies and approximate the excitonic density of states in the full A-L model. We then fit parabolic dispersion curves to the dispersion relations computed in Ref. [75] and we assume that the weighting factors to obtain the average binding energies are the partial density of states corresponding to each exciton (At very low temperatures we might choose the lower exciton values, but we are more interested in intermediate temperatures with $k_B T \gg R_y$). Using this procedure we find that the renormalized masses are $m'_L = 3.51 m_L$ and $m'_h = 0.99 m_h$ This yields a "renormalized" binding energy $R'_y = 1.37 R_y$, a "renormalized" excitonic Bohr radius $a'_B = e^2/(2\varepsilon_0 R'_y)$ and a "renormalized" translational mass $M'_{Lh} = m'_L + m'_h$.

Using our "renormalized" spherical exciton, the calculation of the phonon-induced absorption amounts to making the replacement

$$\frac{1}{V^2}\sum_H\sum_{kq}\delta\left[E_c(k+q) - E_{vH}(k) - \hbar\omega \pm \hbar\Omega_j\right] \to \\ \frac{2}{V}\sum_{nq'}|F_n(0)|^2 \delta\left[E_{\text{ind}} + \frac{\hbar^2 q'^2}{2M'_{Lh}} + E(n) - \hbar\omega \pm \hbar\Omega_j\right] \quad (46)$$

in the expressions for allowed $j =$LA, TO process, such as Eq. (24). Here $q' = q - k_L$, $n$ stands generically for the internal degrees of freedom of the exciton, and $E_n$ is the corresponding energy for the state with an envelope function $F_n(r)$ that satisfies the excitonic Schrödinger equation. For "forbidden" absorption, we face the problem that in Eq. (30) we explicitly considered the anisotropy of the $L$ valley, but we are assuming a spherically symmetric exciton. We must then consider a spherically symmetric deformation potential contribution to keep the problem tractable analytically, replacing Eq. (30) with the simple expression $d'k'$. With the additional wave vector factors, the expression equivalent to Eq. (46) is

$$\frac{1}{V^2}\sum_H\sum_{kq}|k+q-k_L|^2 \delta\left[E_c(k+q) - E_{vH}(k) - \hbar\omega \pm \hbar\Omega_j\right] \\ \to \frac{2}{V}\sum_{nq'}\left|\left.\frac{\partial F_n^*(r)}{\partial r}\right|_{r=0} - is_L q' F_n^*(0)\right|^2 \quad (47) \\ \times \delta\left[E_{\text{ind}} + \frac{\hbar^2 q'^2}{2M'} + E(n) - \hbar\omega \pm \hbar\Omega_j\right]$$

where $s_{L,h} = m_L/(m_L + m_h)$ for near-$L/\Gamma$ processes and $s_{L,h} = m_h/(m_L + m_h)$ for near-$\Gamma/L$ processes. Notice that for hydrogenic excitonic Hamiltonians Eq. (46) implies that only $s$-like functions will contribute to the absorption, a well-known fact from Elliott's theory. For forbidden scattering, on the other hand, the expression contains a derivative of $F_n(r)$, which is only different from zero for $p$-states. This result was already found by Elliott for "forbidden" direct absorption. The difference in the indirect case is that there is also a contribution from $s$-states via the second term in the squared expression. The two terms do not interfere because they are not different from zero simultaneously. Using Eqs. (46) and (47) the excitonic absorption can be computed in a similar fashion as the absorption for free-electron-hole pairs.

*(a) Bound excitons.* The bound exciton absorption for allowed phonon processes can be easily computed using the procedure outlined by Elliott or the approach in Ref. [29] in the constant denominator limit. The result is, for LA phonons:

$$\alpha^\pm_{\text{LA,bound}} = \frac{16\sqrt{2}e^2}{3\pi\rho m_0 c n_{op}}\left(\frac{D^2_{\text{LA}}}{\hbar\Omega_{\text{LA}}}\right)\left(\frac{1}{\hbar\omega}\right)\left(\frac{P^2}{m_0}\right)\frac{(n_{\text{LA}} + \frac{1}{2} \pm \frac{1}{2})}{(E_0 - \hbar\omega)^2} \\ \times \sum_n a'^{-3}_B (M'_{Lh})^{3/2} \frac{1}{n^3}\sqrt{\hbar\omega \mp \hbar\Omega_{\text{LA}} - E_{\text{ind}} + R'_y/n^2} \quad (48)$$

For TO phonons:

$$\alpha^\pm_{\text{TO,bound}} = \frac{64\sqrt{2}e^2}{3\pi\rho m_0 c n_{op}}\left(\frac{D^2_{\text{TO}}}{\hbar\Omega_{\text{TO}}}\right)\left(\frac{1}{\hbar\omega}\right)\left(\frac{\bar{P}^2}{m_0}\right)\frac{(n_{\text{TO}} + \frac{1}{2} \pm \frac{1}{2})}{(\bar{E}_1 - \hbar\omega)^2} \\ \times \sum_n a'^{-3}_B (M'_{Lh})^{3/2} \frac{1}{n^3}\sqrt{\hbar\omega \mp \hbar\Omega_{\text{TO}} - E_{\text{ind}} + R'_y/n^2} \quad (49)$$

For "forbidden absorption" on the other hand, we have separate $p$-state and $s$-state contributions, given by:

$$\alpha^{\pm}_{j\text{-}p,\text{bound}}(\omega) = \frac{16\sqrt{2}e^2}{3\pi m_0 \rho c n_{op}}\left(\frac{d^2}{\hbar^3 \Omega_j}\right)\left(\frac{1}{\hbar\omega}\right)\left(\frac{P^2}{m_0}\right)\frac{(n_j + \frac{1}{2} \pm \frac{1}{2})}{(E_0 - \hbar\omega)^2}$$
$$\times \hbar^2 (M'_{Lh})^{3/2} \sum_n \frac{n^2-1}{3a'^5_B n^5}\sqrt{\hbar\omega \mp \hbar\Omega_j - E_{\text{ind}} + R'_y/n^2} \quad (50)$$

and

$$\alpha^{\pm}_{j\text{-}s,\text{bound}}(\omega) = \frac{32\sqrt{2}e^2}{3\pi \rho m_0 c n_{op}}\left(\frac{d^2}{\hbar^3 \Omega_j}\right)\left(\frac{1}{\hbar\omega}\right)\left(\frac{P^2}{m_0}\right)\frac{(n_j + \frac{1}{2} \pm \frac{1}{2})}{(E_0 - \hbar\omega)^2}$$
$$\times s_L^2 (M'_{Lh})^{3/2} \sum_n \frac{1}{a'^3_B n^3}\left(\hbar\omega \mp \hbar\Omega_j - E_{\text{ind}} + R'_y/n^2\right)^{3/2} \quad (51)$$

for $j = \text{LO, TA}$. Notice that the use of the "renormalized" $M'_{Lh}$ in the above equations corresponds to an enhancement factor smaller than the one used in Eq. (48) of Ref. [29]. We believe that the approach used here is a better approximation of the exact A-L results, but we note that the difference is not very important for our purposes, since at intermediate temperatures the absorption is dominated by the continuum excitonic contribution. Except for the fact that $s_L$ has a different meaning for near-$L/\Gamma$ and near-$\Gamma/L$ "forbidden" processes, the expressions in Eqs. (50)-(51) turn out to be valid in both cases if we use the corresponding linear term coefficient ($d$ or $d\,d'$). This is due to our use of the constant denominator approximation for excitonic effects as well as our approximation of a spherically symmetric deformation potential. As a result of the constant denominator approximation, all of our expressions for bound excitons contain a $(E_0 - \hbar\omega)^{-2}$ divergence. We correct for it in a rather crude way by making the replacement

$$(E_0 - \hbar\omega)^{-2} \rightarrow (E_0 - E_{\text{ind}})^{-2} \quad (52)$$

The justification for this substitution is that bound excitons only play a significant role at the onset of absorption.

*(b) Continuum excitons.* For the continuum solutions, the internal quantum number *n* becomes a wave vector *k*, and the solution can be written as (Ref. [29])

$$F_k(r) = \frac{1}{\sqrt{V}}\Gamma(1+i\nu)e^{\pi\nu/2}e^{i k \cdot r} \,_1F_1(-i\nu, 1, -ikr - i k \cdot r) \quad (53)$$

where $_1F_1$ is a confluent hypergeometric function of the first kind, and $\nu = \sqrt{R'_y/\varepsilon}$, with $\varepsilon = \hbar^2 k^2/(2\mu'_{Lh})$. For allowed absorption we need in Eq. (46) the value

$$|F_k(0)|^2 = \frac{1}{V}|\Gamma(1+i\nu)|^2 e^{\pi\nu} = \frac{1}{V}\frac{\pi\nu}{\sinh(\nu\pi)}e^{\pi\nu} \quad (54)$$

If we insert this into Eq. (46), we notice that in the limit $\nu \to 0$, corresponding to vanishing excitonic interaction, we would expect the two sides to be equal, but since we are using the spherical approximation for excitons, the right hand side would be proportional to $2(m'_L m'_h)^{3/2}$, whereas the left hand side is proportional to $m_\perp m_\parallel^{1/2}(m_{hh}^{3/2} + m_{lh}^{3/2})$, which is about a factor of 2 larger. This is the same problem we faced when we observed that it is not entirely correct to assume separate light-hole and a heavy-hole excitons at the direct gap, except that in that direct case the discrepancy is only 14% (Ref. [29]). To circumvent this problem, we calculate the absorption with the spherical part of the A-L Hamiltonian with and without Coulomb interaction, so that the *ratio* of the two absorptions gives an excitonic enhancement factor $S(\hbar\omega)$ independent of the mass "mismatch". This also bypasses the disadvantages of the constant denominator approximation. We then assume that the experimental continuum absorption can be written as $\alpha^{\pm}_{j,\text{cont}}(\omega) = S^{\pm}_j(\omega)\alpha^{\pm}_{j,\text{free}}(\omega)$. This ensures that in the limit of vanishing Coulomb interaction we recover the free-electron results—including anisotropy—, since in that case $S^{\pm}_j(\omega) = 1$. Furthermore, the diverging factors $(E_0 - \hbar\omega)^{-2}$ from the constant denominator approximation cancel out in the computation of $S(\hbar\omega)$, suggesting that our computed absorption should be much more accurate in the continuum, since we are not forced to make the crude approximation represented by Eq. (52). Carrying out the excitonic calculation, we obtain:

$$S^{\pm}_j(\omega) = \frac{8}{\pi}\frac{\int d\varepsilon \sqrt{\varepsilon}\frac{\pi\nu}{\sinh(\nu\pi)}e^{\pi\nu}\sqrt{\hbar\omega - E_{\text{ind}} \mp \hbar\Omega_j - \varepsilon}}{(\hbar\omega - E_{\text{ind}} \mp \hbar\Omega_j)^2} \quad (55)$$

for $j = \text{LA, TO}$. An example of this function is shown in Fig. 6. For "forbidden" absorption, we need the derivative of the wave function in Eq. (53). This derivative contains two terms, of which the second one is proportional to $\nu$ and thus smaller. Neglecting this contribution we then obtain

$$\left.\frac{\partial F^*_n(r)}{\partial r}\right|_{r=0} - is_L q' F^*_n(0) = \frac{i}{\sqrt{V}}\Gamma(1+i\nu)e^{\pi\nu/2}[k - s_L q'] \quad (56)$$

When squaring this expression there is a term proportional to $k \cdot q'$ that vanishes upon angular integration, and we finally obtain an excitonic enhancement function

$$S^{\pm}_j(\omega) = \frac{16}{\pi}\frac{\int d\varepsilon \varepsilon^{3/2}\frac{\pi\nu}{\sinh(\nu\pi)}e^{\pi\nu}\sqrt{\hbar\omega - E_{\text{ind}} \mp \hbar\Omega_j - \varepsilon}}{(\hbar\omega - E_{\text{ind}} \mp \hbar\Omega_j)^3} \quad (57)$$

for $j = \text{LO, TA}$. An example of this function is also shown in Fig. 6.

TABLE I. Electronic and band structure parameters for Ge used in the computation of PL spectra. Values are given for room temperature ($T = 295$ K). The second row indicates how the temperature dependence of the parameters was accounted for. If the temperature dependence was neglected, a dash appears in this second row. Some parameters are not listed because they can be computed from the table entries using formulas given in the text.

| Temperature | $E_{ind}$ (eV) | $E_0$ (eV) | $E_1$ (eV) | $\Delta_0$ (eV) | $\Delta_1$ (eV) | $P^2/m_0$ (eV) | $\bar{P}^2/m_0$ (eV) | $\Delta_L$ (eV) | $\frac{m_\parallel}{m_0}$ | $\frac{m_{hh}}{m_0}$ | $\frac{m_{lh}}{m_0}$ |
|---|---|---|---|---|---|---|---|---|---|---|---|
| 295 K | 0.660 | 0.805 | 2.109 | 0.287[a] | 0.200[b] | 12.61 | 12.94 | 1.405 | 1.58[c] | 0.352 | 0.0386 |
| 0-295 K | See text | Ref. 29 | Ref. 29 | — | — | Ref. 29 | Ref. 29 | Ref. 47 | — | Ref. 29 | Ref. 29 |

[a]Ref. [77], at $T = 10$K
[b]Ref. [47]
[c]Ref. [78]

## V. EXPERIMENTAL DATA FITS

The calculation of the PL emitted from a point at depth $z$ below the sample surface begins with an estimate of the photoexcited carrier concentration via Eq. (8), which is needed to determine the quasi-Fermi levels using Eq. (9) and (11). These quasi-Fermi levels are then used to calculate the absorption. We start with Eqs. (26), (28), (32), and (35), modified to account for broadening as discussed in Sec IV.G. We then multiply each of these expressions times the corresponding excitonic enhancement factor in Eq. (55) or (57) to obtain the excitonic continuum contribution to the absorption. Next we add the bound exciton contribution for each phonon mode using Eqs. (48), (49), (50) and (51). with the approximation in Eq. (52). An additional approximation is that we use for bound excitons the same anisotropic deformation potential parameter $d'_{LO}$ used for the continuum exciton.

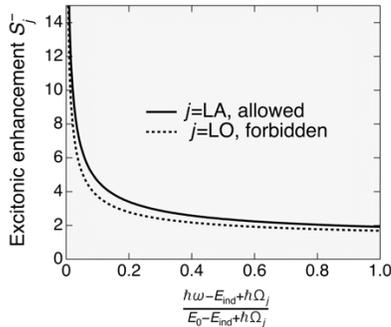

FIG. 6. Excitonic continuum enhancements for indirect absorption with phonon annihilation (which maps into indirect PL with phonon emission) for the allowed and forbidden cases. The horizontal scale is selected so that zero corresponds to the absorption edge and 1 to the direct gap.

The phonon-assisted absorption calculations are carried out for the cases of phonon-creation and annihilation and added up. Next we add the direct absorption contribution from Eq. (40). The final step is to compute the PL spectrum by inserting the total absorption into Eq. (1). The spectrum is corrected for reabsorption of the emitted radiation by multiplying times $\exp[-\alpha(\omega)z]$. We also account for final state broadening by convolving the calculated spectrum with a Lorentzian and or Gaussian. Since the spectra at intermediate temperatures are rather broad, this convolution has negligible impact on the computed PL except at the lowest temperatures.

The PL calculation is repeated at several depths and averaged. We use a total of 21 points, separated by 700 nm near the surface and increasing the separation as a function of depth. Depth sampling could be avoided in thin film measurements, as discussed below.

The input material data in the calculation are from Table I and II, which leaves as the sole adjustable parameters the deformation potentials $D_{LA}$ and $D_{TO}$ and the deformation potential derivatives $d'_{LO}$ and $d_{TA}$ (the assumption $d'_{TA} = 0$ is discussed in more detail in Section VII.). These four parameters are expected to reproduce the temperature-dependent PL lineshapes in Fig. 2. This includes a correct ratio of indirect/direct PL luminescence, since the latter does not depend on deformation potentials. In addition, we expect the calculated indirect absorption with these parameters to match the experimental indirect absorption in Refs. [28,29], the experimental broadening of the direct gap exciton measured by Li et al. [43], and the width of the intermediate states calculated by Tandon et al. [40], the latter two conditions via Eq. (38). These multiple requirements could in principle be implemented in a least-squares fitting routing that considers all experimental and theoretical information at once. However, such procedure is numerically difficult and will not necessarily lead to an optimal choice of deformation potentials, since the approximations we made to obtain analytical expressions have different degrees of validity depending of the photon frequency $\hbar\omega$.

TABLE II. $L$-point phonon frequencies used in the computation of PL spectra. The values correspond to $T = 80$K but were used unchanged at all temperatures. See text for discussion.

| $\Omega_{TA}$ (meV) | $\Omega_{LA}$ (meV) | $\Omega_{LO}$ (meV) | $\Omega_{TO}$ (meV) |
|---|---|---|---|
| 7.86[b] | 27.5[b] | 30.4[b] | 36.0[b] |

[b]Ref. [79]

TABLE III. Deformation potential parameters fit to the experimental data compared with theoretical values. The errors are those from the PL fit at 160 K and do not include the uncertainty from the many approximations in the theoretical PL model.

|  | $D_{LA}$ ($10^8$ eV/cm) | $D_{TO}$ ($10^8$ eV/cm) | $d_{TA}$ (eV) | $d'_{LO}$ (eV) |
|---|---|---|---|---|
| Exp. | 1.31±0.04 | 3.18±0.65 | 45±1 | 69±3 |
| Tandon *et al.* [40] | 1.58 | 1.76 | 10.0[a] | 12.2 |

[a] Along the <110> direction.

We then adopt a much simpler procedure that leads to satisfactory results. We start by fitting the experimental spectrum at 160 K. From this fit we obtain *relative* values of the deformation potentials. The only weak contribution at this temperature turns out to be the one from the TO phonon, which we further adjust by fitting the TO/LA ratios at 12 K. The absolute values are next adjusted by matching the indirect absorption from Refs. [28,29]. No attempt is made to further adjust the deformation potentials to match the experimental indirect/direct PL intensity ratios or the experimental and theoretical broadenings. These quantities are simply compared with the predictions as discussed below.

## VI. RESULTS

Our deformation potential parameters are listed in Table III, where the errors correspond to the 160 K fit. These parameters lead to excellent agreement with the PL lineshapes at all temperatures, as seen in Fig. 2, suggesting that our model captures the most important contributions to indirect gap PL. Figure 7 shows the calculated absorption using the same deformation potentials from Table III, and we see that the agreement is also very good, providing a self-consistent confirmation of the model. In Fig. 8 we show the calculated width of the conduction band states near the $\Gamma$-point using Eq. (38) with the deformation potentials from Table III and compare them with the *ab initio* results from Tandon *et al* [40] and with experiment from Ref. [43].
Finally, we notice that the deformation potential choice in Table III leads to good agreement with direct/indirect PL intensity ratios in Fig. 2. These ratios are difficult to quantify because our predicted lineshapes show their maximum deviation from experiment at the onset of direct-gap PL. Nevertheless, it is apparent from Fig. 2 that the relative strength and its temperature dependence are correctly predicted.

Our theoretical expressions also make it possible to evaluate the integrated intensity of the PL signal as a function of temperature, and the result is shown as a solid red line in Fig. 3. The agreement with experiment is not good, but we note that the discrepancy is mostly due to the uncertainty in the temperature dependence of the ambipolar diffusivity, as discussed in Section III. Our use of the experimental temperature dependence of the carrier mobilities leads to a diffusivity that increases too fast as the temperature is decreased. This lowers the photo-excited carrier concentrations and reduces the predicted integrated intensity, in disagreement with experiment. The dashed red line shows the same calculation but keeping the diffusivity fixed at the room temperature value, and we see that theory now predicts a very rapid increase of the integrated intensity as the temperature is lowered, far exceeding the observed increase. The experimental values are between those two extremes.

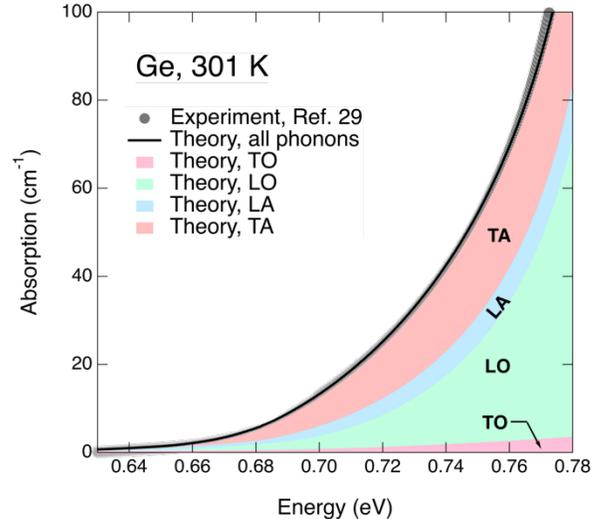

FIG. 7. Experimental indirect absorption in Ge, from Ref. 29 and calculated absorption obtained by adding up the contributions from TA, LA, LO, and TO phonons with deformation potentials from Table III, so that the relative strength of these contributions are consistent with the fit of the PL spectra. The color-shaded areas show the individual contributions.

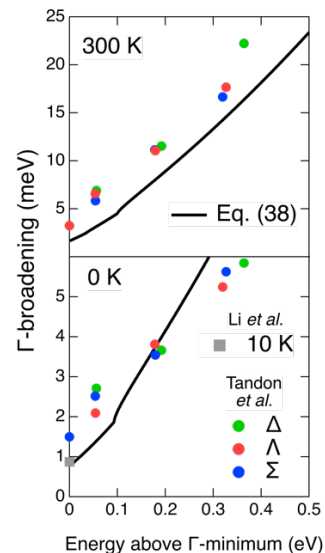

FIG. 8. Half-width (imaginary part of the self-energy) for states near the $\Gamma$-minimum of the conduction band. The solid line is from Eq. (38) using experimental deformation potentials from Table III. Circles correspond to *ab initio* calculations from Refs. [40],[80], with colors identifying the different directions away from the $\Gamma$-point in the *ab initio* calculation. The square dot is the experimental value from Ref. [43].

# VII. DISCUSSION

### 1. Choice of material parameters

The predicted PL lineshapes in Fig. 2 are generally rather insensitive to the precise value of the material parameters in Table I and II. The only exceptions are the temperature dependence of the indirect band gap and the phonon frequencies, which we discuss in this section.

Most of the literature references on the temperature dependence of the indirect gap point to the classic 1960 review by McLean [37]. This author measured optical absorption and extracted the band gap energies by inspection of the absorption edges. He used a theoretical model which close to the absorption edge has an energy dependence similar to the model presented here. The exact procedure used to fit the theory to experiment seems to have been a combination of visual inspection and numerical methods. An obvious difficulty of this approach is that, due to lifetime broadening, the identification of the absorption edges becomes harder as the temperature is raised. This is a clear disadvantage of absorption measurements when compared to PL measurements. It becomes apparent by comparing Fig. 6 with the PL spectrum at 295K in Fig 2. Whereas the former is a smooth curve, the latter has a well-defined edge and a clear peak.

The data from McLean were processed in another classic paper by Varshni [81], who introduced his well-known formula $E_{ind}(T) = E_{ind}(0) - \alpha T^2/(\beta + T)$ to represent the temperature dependence. The parameters of the fit were found to be $E_{ind}(0) = 0.7412$ eV, $\alpha = 4.561 \times 10^{-4}$ eV/K, and $\beta = 210$ K. These parameters have been widely used in the literature to represent the temperature dependence of the indirect gap. However, a reevaluation of the same data by Thurmond [82] gives $E_{ind}(0) = 0.7437$ eV, $\alpha = 4.774 \times 10^{-4}$ eV/K, and $\beta = 235$ K. We have carried out our own fit of the McLean data and we find $E_{ind}(0) = 0.7446$ eV, $\alpha = 4.777 \times 10^{-4}$ eV/K, and $\beta = 231$ K, in much better agreement with Thurmond. Using this band gap, however, we find excellent agreement between the PL lineshapes and experiment at low temperature but a rigid shift that increases with temperature to reach about 5 meV at room temperature. This is small enough for our fits to be very close to the experimental data, but since we believe that this shift is caused by the difficulties in determining absorption edges near room temperature using the McLean method, we have changed the Varshni parameters to $E_{ind}(0) = 0.7440$ eV, $\alpha = 4.956 \times 10^{-4}$ eV/K, and $\beta = 217$ K. This choice leads to the nearly perfect energy match between the main features of the theoretical and experimental PL curves at all temperatures.

The phonon energies used in our simulations are those in Table II, which were measured at 80 K [79]. The shifts between 12 K and 295 K should be less than 0.5 meV [83], which is almost negligible in the scale of our measurements but could have been included for completeness. However, we have not considered this correction because our model neglects a larger effect, namely the wave-vector dependence of the phonon frequencies. While this dependence will not have any significant impact on the overall PL lineshape either, it could be responsible for some of the "high-frequency" features in the PL spectra. The neglect of the wave vector dependence of the phonon energies is obviously a worse approximation for "forbidden" processes, since in those cases the phonons with a wave vector exactly at the $L$ point will have a vanishing electron-phonon coupling. We expect the largest effect for TA-phonons, both because their frequency is very low and because of strong dispersion in directions perpendicular to <111>. In fact, it is apparent in the 50 K spectrum in Fig 2 that the predicted TA peak is somewhat shifted relative to its experimental counterpart. If we change the phonon frequency to match the PL peaks exactly, we notice that one important side-effect is a change in the value of deformation potential derivative $d_{TA}$ that best fits the data. This is because of the presence of the inverse phonon frequency as a prefactor in all absorption expressions.

### 2. Failures of the PL model

In spite of the remarkable agreement between theory and experiment in Fig. 2, there are a few but significant remaining discrepancies. Most notable are the failure of theory to reproduce the intensity of the TA peak at the lowest temperature of 12 K, and the failure to reproduce the experimental lineshape just below the peak associated with direct gap emission, particularly near room temperatures. The intensity of the theoretical TA peak at 12 K is roughly 3 times weaker than observed. This is actually not entirely surprising given the crudeness of our bound exciton model, as explained above. We are using two degenerate hydrogenic excitons to model two non-degenerate, highly non-parabolic excitons that display a "mass reversal" effect [75]. The approximation is likely to be even worse in the case of "forbidden" processes, which in the hydrogenic limit involve non-interfering $s$-state and $p$-state contributions. The real excitons in Ge lack spherical symmetry and the relevant matrix elements for bound excitons may be very different. Furthermore, while our excitonic model for direct transitions accounts for screening by the photoexcited carriers, this is not the case for our indirect excitons. As discussed above, bound excitons are most sensitive by this screening, and this may contribute to the observed discrepancy if screening affects allowed and "forbidden" transitions differently.

An alternative explanation for the failure to reproduce the observed strength of low-temperature TA-assisted emission is the possible existence of an anharmonic decay bottleneck for TA phonons [84]. At low temperatures, the PL intensities are extremely sensitive to the value of the vanishing phonon-occupation numbers, and even a minor deviation from the thermal values might be enough to change the predicted intensities appreciably.

The lineshape discrepancies near the direct band gap $E_0$, which become quite apparent at room temperature, are difficult to unravel because at least three factors may contribute: (a) there is a sharp cutoff of the indirect emission at this energy as a result of self-absorption, but this is obtained from the quasi-1D illumination model discussed in Sect. III. A more realistic 3D model could broaden this edge; (b) the intensity enhancement due to the excitonic nature of the intermediate states, which we are neglecting to keep the model tractable, peaks at these energies; and (c) the resonant $E_0$-route indirect processes diverge at this energy. We are using approximate expressions to calculate the indirect contributions in this range, and it is not even obvious if second-order perturbation theory is applicable in this regime. The treatment of the near-$E_0$ spectral region would be highly simplified in measurements on thin films, as discussed below, and this may allow us to improve our model at the borderline between mainly indirect and mainly direct emission.

### 3. Allowed vs "forbidden" absorption

Our model implies that indirect absorption and emission near room temperature are dominated by "forbidden" TA and LO processes. LA phonons, which were assumed to completely dominate room temperature indirect absorption in Refs. [28] and [29], make a relatively small contribution in Fig. 6, and the fit parameters in Table III also show a strong enhancement of the "forbidden" channels relative to the theoretical predictions. This is not entirely surprising given previous evidence—for example in the case of GaAs [85,86]—, that "forbidden" TA phonons make a substantial contribution to intervalley relaxation at room temperature. To better understand this result, it is very useful to begin by analyzing the contributions to the PL signal from each individual phonon type. This is done in Fig. 9, which shows the PL associated to TA, LA, LO, and TO phonons at three selected temperatures.

Each phonon contribution consists of two peaks: a lower-energy one which is the reverse of photon absorption with phonon annihilation, and a higher-energy peak which corresponds to the reversal of photon absorption with phonon creation. Each contribution is associated with the thermal factors $n_j$ and $(n_j + 1)$, respectively, but the *former* dominates at low temperature due to the exponential in the denominator of Eq. (1). Both peaks and their temperature dependence are clearly seen in each of the colored traces in Fig. 9, except for the TA line, for which the vibrational frequency is very low and the peaks are mostly merged. The high-energy peak is in all cases *closer* to the direct gap $E_0$. This is important for phonon processes via the $E_0$-route, because it produces a sizable enhancement of its relative contribution due to the resonant character of this route. This is apparent by comparing the two contributions in the case of LA phonons ($E_0$-route) and TO phonons, ($E_1$-route).

In the case of "forbidden" processes, the frequency dependence near the absorption threshold is qualitatively proportional (ignoring denominator and excitonic effects) to $\left(\hbar\omega \mp \hbar\Omega - E_{\text{ind}}\right)^3$, whereas the allowed processes are proportional to $\left(\hbar\omega \mp \hbar\Omega - E_{\text{ind}}\right)^2$. This implies a more rapid rise of the "forbidden" contributions, which combined with the $E_0$-route resonance lead to high-energy tails that may extend all the way to and past the direct gap, as seen for the LO case at room temperature in Fig. 9. On the other hand, at the onset of absorption the cubic energy dependence of "forbidden" processes makes them weaker than allowed ones. This is amplified by the fact that the excitonic enhancements from Eqs. (55) and (57) are much stronger near the absorption edge [29]. This is also apparent in Fig. 9, where the signal rises very sharply on the low-energy side due to the excitonic enhancement of the LA and TO contributions, but the rise of the forbidden TA and LO signals has a much lower slope.

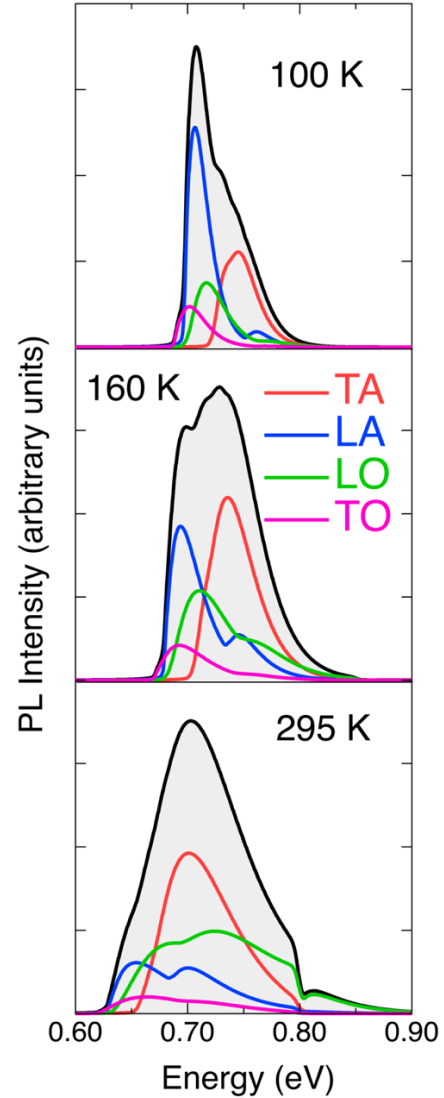

FIG. 9. Decomposition of the theoretical indirect PL lineshape for Ge into separate phonon contributions. The solid black line is the overall indirect-gap contribution to the PL signal that appears in Fig. 2. The colored lines show the PL associated with TA (red), LA (blue), LO (green) and TO (Magenta) phonons.

The above considerations and the phonon decompositions in Fig. 9 allow us to draw important conclusions: (a) "forbidden" processes are relatively weaker at low temperatures because their associated absorption near the absorption edge is lower, and the exponential in the denominator of Eq. (1) suppresses contributions far away from the edge. As the temperature is raised, however, energies away from the edge begin to make a contribution, and since the "forbidden" absorption increases "faster" than the allowed absorption, its relative weight increases. The emergence of "forbidden" emission in Fig. 9 as a function of temperature is very clear. This explains why "forbidden" processes were far less important for previous studies, which were mostly limited to cryogenic temperatures; (b) TA phonons are essential to explain the PL lineshape and its temperature dependence. The phonon frequencies are quite similar for LA, TO, and LO phonons, and it is the low-frequency TA modes with their much stronger temperature dependence which causes the changing lineshape. In particular, the relatively sharp peak at room temperature is essentially due to TA phonons. Our model then captures the relative strength of TA phonon emission correctly, otherwise it would be impossible to match the overall PL lineshape. But this implies that TA phonons must make an important contribution to the room-temperature absorption, as illustrated in Fig. 7; (c) we noticed in the theory section that there are several possible TA phonon contributions to the absorption. In particular, there is a "near-$L/\Gamma$" channel—akin to the LO process in Fig. 9—that was neglected because it would also contribute to the relaxation of $\Gamma$-point electrons to the $L$-valley, a process in which TA phonons have a negligible participation, according to *ab initio* simulations by Tyuterev *et al.* [39]. We now see that if we were to include this process to represent the TA phonon contribution to the PL, the predicted lineshape would be too broad on the high-energy side because "near-$L/\Gamma$" processes have a $\left(E_0 - \hbar\omega\right)^{-1/2}$ divergence. This is clearly apparent for the LO-phonon contribution (green trace) in Fig. 9. Instead, in "near-$\Gamma/L$" processes the divergence is suppressed. This leads to the correct lineshape while at the same time not contributing to the relaxation of $\Gamma$-point electrons to the $L$-valley, as is evident from the fourth term in Eq. (38). Thus, the TA mechanism represented by Eq. (35) is crucial to insure agreement between experiment and the predicted lineshapes near room temperature.

While the evidence for a TA contribution is overwhelming, and the specific mechanism in Eq. (38) is the only one that reproduces the experimental lineshape, the case for LO phonons is more subtle. A distinct LO signal appears as a weak shoulder in high-resolution PL experiments at 4.3 K [31], but otherwise there are no features that can be unambiguously assigned to an LO contribution. If we remove LO phonons from our model, we can still obtain a reasonably good agreement with the overall temperature dependence of the PL lineshapes. The contribution from the other phonons can be increased to match the experimental absorption, but the nearly perfect agreement seen in Fig. 7 is no longer obtained, as the theoretical curve no longer matches the slope of the experimental data. Furthermore, the predicted width calculations are worsened relative to Fig. 8. But we cannot completely rule out the possibility that our model is artificially enhancing the LO phonon strength to mimic excitonic effects not included in our simplified treatment. We note that in the 295 K spectrum in Fig. 9, the LO signal is the strongest one at the direct gap threshold. As discussed before [28,29], the constant denominator approximation exaggerates the excitonic enhancement at the onset of absorption and misses the enhancement near the direct gap due to the excitonic nature of the intermediate states. Thus the lineshape of the PL signal near the direct gap and the lineshape of the absorption curve, might still be matched with a smaller LO contribution using a more realistic treatment of excitonic effects. We discuss below ways how this could be tested.

A comparison of the experimental and theoretical deformation potentials in Table III shows that for allowed processes the agreement is quite good. However, $D_{\text{LA}}$ turns out to be considerably lower than previously accepted. The literature value $D_{\text{LA}} = 4.2 \times 10^8$ eV/cm seemed particularly robust because it was obtained from time-resolved transmission [42], broadening of the direct gap exciton [43], and absorption [28,29], and confirmed by the theoretical results from Tyuterev *et al.*[39], Krishnamurthy and Cardona [87], and Murphy-Armando and Fahy[88]. The new value in Table III, however, is still consistent with all available experimental measurements if we include "forbidden" processes, although the reasons for the discrepancies between theoretical calculations are not clear to us. Table III also shows a discrepancy of a factor of almost 2 in $D_{\text{TO}}$ between theory and experiment, but the experimental value was adjusted to match the relative TO/LA intensity ratios at 12 K. This may not be warranted because our model for bound excitons is a gross oversimplification, as discussed earlier. At higher temperatures, the TO contribution is modest, and reducing $D_{TO}$ to bring it closer to the theoretical prediction would hardly make an impact on the predicted PL lineshapes. On the other hand, the deformation potential derivatives for "forbidden" processes fit to the experimental data are at least a factor of 4 higher than the theoretical ones, (although the relative LO/TA magnitude is in much better agreement with theory). The enhancement of the "forbidden" contribution relative to theoretical predictions can also be seen if we compare the predictions of our model with the calculations from Tyuterev et al.[39]. They find that for the Γ-minimum in the CB, 36% of the width originates from LO phonons and 64% from LA phonons, whereas our calculation in Fig. 8 with the experimental deformation potentials from Table III is composed of LO and LA fractions of 92% and 8%, respectively.

The discrepancies between different theoretical values for deformation potentials make it hard to compare experiment to theory. More fundamentally, the experimental deformation potentials may not be directly comparable to theoretical ones and should be viewed as "effective" values. For example, we note that Tandon et al [40] find that $D_{LA}$ has a maximum value when it couples the Γ and L points exactly, whereas our fit value represents an average coupling. Taking this into account, the agreement between theory and experiment for $D_{LA}$ may be even better than shown in Table III. Furthermore, we have intentionally neglected several absorption routes, which, when combined, could make a sizable contribution to the overall PL and end up being treated "effectively" by our model. Some indirect evidence that this is the case is provided by the observation that recalculating the widths in Fig. 8 using Eq. (38) but with the Tandon et al. parameters from Table III instead of the experimental ones, we actually *worsen* the agreement with the widths computed by Tandon et al by adding over all possible decay channels. Ultimately, the best way to compare theory and experiment would be via predictions of the PL spectrum itself, adding numerically over all possible channels.

### 4. Further improvements

The motivation for carrying out the PL measurements in a bulk Ge sample was the need to minimize any possible contribution from no-phonon transitions, which had been previously invoked to explain the PL lineshape. However, the very high ambipolar diffusivity in Ge creates serious modeling challenges, since depth sampling becomes necessary, the details of the lineshape may depend on the illumination geometry, and the integrated intensity is strongly diffusion-dependent. Measurements of the ambipolar diffusivity as a function of temperature are needed to resolve the order-of-magnitude difference between the excitonic diffusion at low temperatures and the predictions based on the temperature dependence of carrier mobilities,

Indeed, the ultimate solution to bypass these complications would be to move away from bulk Ge and carry out measurements on Ge films with thicknesses below the diffusion length. Our demonstration in this work that the PL from bulk Ge can be accounted for quantitatively without the need to invoke no-phonon transitions implies that the same should be true in Ge films, provided that the density of defects can be kept sufficiently low. It remains to be seen if this is the case for the ubiquitous Ge-on-Si films, grown either by Molecular Beam Epitaxy [89], the standard two-step Chemical Vapor Deposition method [90], or using modern low-temperature chemistries [91]. If dislocation densities in such films are too high to suppress no-phonon lines, this growth would have to carried out on nearly lattice-matched substrates such as GaAs.

The photoexcited carrier density in a Ge film with thickness below the ambipolar diffusion length is expected to be very uniform, so that the calculations could be performed at a single point. This would dramatically reduce the computation time, allowing us to carry out the integrals containing broadening numerically, and to use the more realistic excitonic model of Refs. [28,29].

Further model improvements to better capture the basic physics of the indirect PL process in Ge can be made with assistance from theory. Full blown microscopic calculations such as those performed in Si [11] are much more challenging in Ge due to the smaller band gap, the strong spin-orbit interaction, and the sensitivity to the energy difference between direct and indirect gaps. The need to include excitonic effects further complicates the theoretical challenge. However, microscopic calculations of the electron-phonon coupling such as those performed in Ref. [40] could be used to quantify the importance of the different phonon channels and absorption routes, and to extract effective, energy dependent phonon frequencies that could be incorporated in the model. At the direct gap threshold, the entire perturbation theory approach could collapse due to the divergence of the energy denominators. To analyze this regime, non-perturbative calculations would be highly desirable, and for this purpose the elegant method introduced by Zacharias et al. [12] may prove extremely valuable by making it possible to estimate the contribution from phonon-assisted processes at the direct gap and above.

## VIII. CONCLUSIONS

We have presented detailed PL measurements from high-quality bulk Ge samples as a function of temperature. A theoretical PL model based on the fundamental van Roosbroeck-Shockley (RS) equation has been developed and shown to be in remarkable agreement with the experimental measurements. The model consists of analytical expressions that can be used to fit experimental data, and should prove useful for the spectroscopy of Ge-like materials such as $Ge_{1-x}Si_x$, $Ge_{1-y}Sn_y$, and related ternary compounds.

Our model fits indicate that "forbidden" indirect absorption processes play a dominant role near room temperature. In particular, a TA phonon contribution determines the characteristic lineshape of the room temperature PL, and explains the strong temperature dependence of this lineshape.

## ACKNOWLEDGEMENTS

Numerous discussions and exchanges of information with Dr. Nandan Tandon and Dr. Pengke Li are gratefully acknowledged. In particular, Dr. Tandon not only shared with us numerical values of his calculations in Ref. [40], which we used to generate Fig. 7, but contributed additional data points calculated expressly for this figure. Illuminating discussions with Dr. Li helped us compute the electron-phonon matrix elements needed for the modeling of TO-assisted PL. Additional discussions with Drs. Eugenijus Gaubas, Enrico Belloti, and Stefano Dominici helped us build a quantitative model of non-radiative recombination.

This work was partially funded by the AFOSR under grant FA9550-17-1-0314.